\documentclass[%
 preprint,
 amsmath,amssymb,
 aps,
 prc,
 floatfix,
 nofootinbib
]{revtex4-2}

\usepackage{graphicx} 
\usepackage{xcolor}
\usepackage{hyperref}

\newcommand{\keyword}[1]{{\bf #1}}

\begin{document}

\title{What we talk about when we talk about nuclear structure}

\author{Ragnar Stroberg}
\date{\today}

\begin{abstract}
I provide and introductory overview of the field of nuclear structure, with a focus on physical concepts.
I describe some basic nuclear structure observables, followed by a qualitative description of nuclear forces.
I then outline some nuclear structure models which are most widely used to interpret the experimental data in terms of interacting protons and neutrons.
\end{abstract}

\maketitle


\section{Introduction}
The term ``nuclear structure'' reflects the notion that it is often useful to think of the atomic nucleus as made up of smaller pieces.
The most commonly used pieces are protons and neutrons (collectively called \keyword{nucleons}),
but it can also be useful to think of a nucleus as made of clusters like alpha particles, or as a continuous nuclear ``liquid''.
At a more fundamental level, we may think of the nucleus as made of quarks.
The study of nuclear structure largely consists in understanding how the properties and behavior of these sub-nuclear degrees of freedom manifest as observable properties of the nucleus.
Stated more simply, we are interested in \emph{what} the nucleons are doing inside the nucleus, and \emph{why} they are doing it. 

We begin in section~\ref{sec:Symmetries} by briefly reviewing the symmetries relevant for nuclear structure; these provide conserved (or approximately conserved) quantum numbers we use to label states.
In section~\ref{sec:Observables} we review the most commonly considered nuclear structure observables; these provide the data against which any model of the nucleus should be compared.
In section~\ref{sec:NNforce} we describe our understanding of the force between nucleons; this provides the essential building block for understanding what nucleons do in nuclei.
In section~\ref{sec:Phenomena} we describe several important phenomena in nuclear structure, and the conceptual frameworks used interpret the experimental data.
Finally, in section~\ref{sec:StructureToConstrainForces}, we briefly discuss ways in which nuclear structure observables can help inform our knowledge of the nuclear force.

Essentially no attempt is made to describe the technical details of experiments, or theoretical formalisms.
The focus is on the conceptual framework.
Important technical jargon is highlighted in bold font upon first use.
References to more detailed accounts are provided for the interested reader.


\section{\label{sec:Symmetries}Symmetries and quantum numbers}
It is helpful to consider symmetries (or approximate symmetries) of the interactions between nucleons, because these provide conserved quantum numbers with which to label nuclear states, and selection rules for transitions between states.
The most relevant symmetries for nuclear structure are rotational invariance, parity, and isospin.

Rotational invariance reflects the fact that there is no preferred spatial direction, and it leads to the conservation of total \keyword{angular momentum}, which we label with $J$.
This angular momentum is constructed from the spin and orbital angular momentum of each of the nucleons in the nucleus.
Because nucleons have spin $\tfrac{1}{2}$, and orbital angular momentum takes integer values of $\hbar$, a nucleus with an even number of nucleons will have integer $J$, while an odd number of nucleons will have half-integer $J$.

Strong and electromagnetic interactions are invariant under a \keyword{parity} transformation $(x,y,z)\to (-x,-y,-z)$, corresponding to switching from right-handed to left-handed coordinates.
Two applications of the parity transformation return to the original coordinates, so under parity a wave function can at most change by a sign: $\psi(-x,-y,-z)= \pm \psi(x,y,z)$.
Nuclear states are labeled by the parity quantum number $\pi=\pm 1$.
Weak interactions are not invariant under parity transformations~\cite{WuPR1957} so parity is technically an approximate symmetry in nuclei, but the effects of parity mixing are generally negligible due to the weakness of the weak interaction.

Finally, due to the lightness of the $u$ and $d$ quark masses and the weakeness of electromagnetism~\cite{BedaqueARNPS2002}, strong nuclear interactions are essentially the same for proton-proton, neutron-neutron, or proton-neutron pairs, after accounting for the Pauli principle.
Likewise, the proton and neutron masses differ by $\sim 0.1$\%.
This can be incorporated by treating protons and neutrons as the same type of particle (the nucleon) with an additional internal degree of freedom called \keyword{isospin}, which follows the mathematical structure of angular momentum.
The nucleon has isospin $t=\tfrac{1}{2}$, with the two projections $t_z$ corresponding to the proton and the neutron.
In the particle physics convention, protons have $t_z=+\tfrac{1}{2}$ ``isospin up'') while neutrons have $t_z=-\tfrac{1}{2}$ ``isospin down''.
In the nuclear physics convention, this is reversed.
Just as the individual angular momenta of all the nucleons combine to a total angular momentum $J$ and projection $M$, the isospins of all the nucleons in a nucleus combine to a total isospin $T$ and projection $T_z$.
The isospin projection is simply $T_z=\tfrac{1}{2}(N-Z)$ (in the nuclear physics convention), while the total isospin $T$ encodes more subtle properties of the wave function.
Of course, the Coulomb interaction only affects proton-proton pairs, and so isospin symmetry is broken by electromagnetic interactions.
It is an approximate symmetry, whose breaking becomes more severe with increasing proton number $Z$.

\begin{figure}
    \centering
    \includegraphics[width=0.75\linewidth]{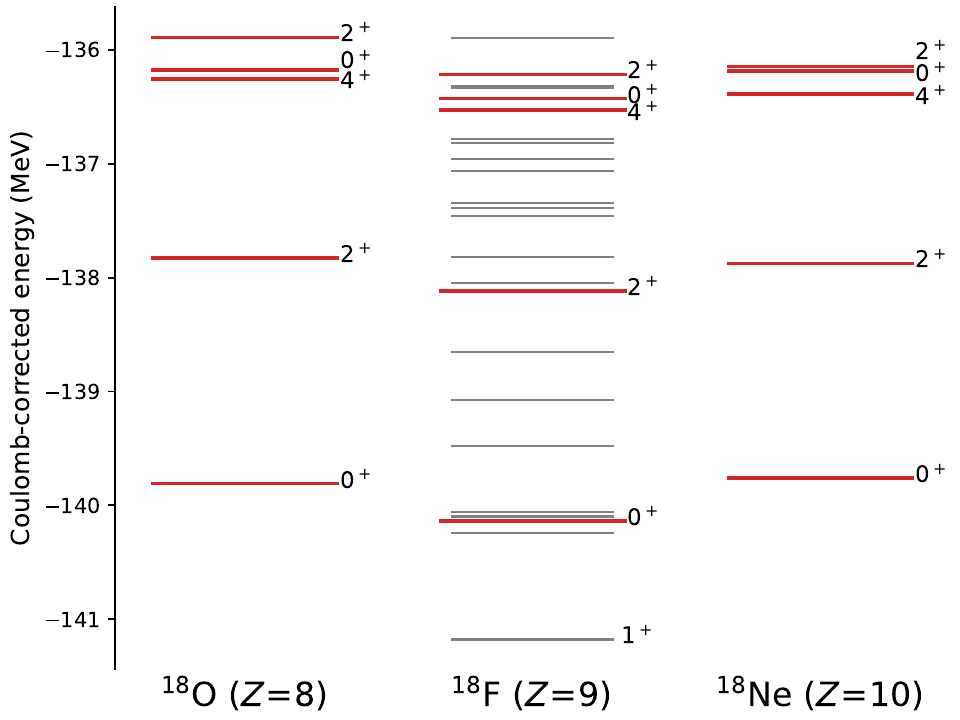}
    \caption{Spectrum of states in $A=18$ nuclei, highlighting the isobaric analogue states in red. The overall energies have been corrected for the Coulomb interaction. The labels to the right of the levels indicate the angular momentum and parity $J^\pi$. Experimental data from the NNDC~\cite{NNDC}.}
    \label{fig:IAS}
\end{figure}

One striking consequence of isospin symmetry is the existence of \keyword{isobaric analogue states}.
For a total angular momentum $J$, there are $2J+1$ degenerate states corresponding to different projections $M=-J\ldots +J$.
Likewise, for total isospin $T$, there will be $2T+1$ approximately degenerate states corresponding to different projections $T_z=-T\ldots +T$, which correspond to different combinations of protons and neutrons with the same total number of nucleons.
Figure~\ref{fig:IAS} shows the isobaric analogue states in the $A=18$ isotopes $^{18}$O, $^{18}$F, $^{18}$Ne.
To more clearly compare the spectra, the effects of the Coulomb interaction have been modeled as two particles outside a $^{16}$O core through the formula
$
\Delta E_{\rm Coul} = \frac{e^2 8(Z-8)}{R}
$
where the surface radius is taken to be $R=1.2 A^{1/3}$~fm (see section~\ref{sec:DensitiesRadii}).

 Other, even more approximate, symmetries are occasionally discussed in various contexts, including Wigner's spin-isospin $\mathrm{SU}(4)$ symmetry~\cite{Wigner1937} and Elliot's $\mathrm{SU}(3)$ symmetry of the 3D harmonic oscillator~\cite{ElliotPRSA1958I}.

\section{\label{sec:Observables}Nuclear structure observables}

In this section we discuss some of the most commonly used observables for probing nuclear structure, and describe some ways in which they can be measured.
A vast amount of nuclear structure data can be found on the web page of the National Nuclear Data Center~\cite{NNDC}

\subsection{Mass and binding energy}
Perhaps the most fundamental observable property of a nucleus is its mass.
To a good approximation, the mass of a nucleus with $Z$ protons and $N$ neutrons is simply the sum of the masses of the constituent nucleons.
However, because the nucleons are bound together their energy is less than it would be if they were separated, and by Einstein's $E=mc^2$ the mass is slightly reduced:
\begin{equation}
M(Z,N) = Z m_p + N m_n - BE(Z,N)/c^2
\end{equation}
where $m_p$ and $m_n$ are the proton and neutron masses, respectively, and $BE(Z,N)$ is the \keyword{binding energy} due to the motion of the nucleons and their interactions.
Equivalently, masses can be expressed in terms of the ``mass excess''
\begin{equation}
    M_{\rm ex}(Z,N) = M(Z,N) - A\cdot u
\end{equation}
where $A=Z+N$ and $u\approx 931.494$~MeV/$c^2$ is the atomic mass unit, defined as $1/12$ of the mass of $^{12}$C.

Measurement of the mass provides a direct measure of the total energy (kinetic plus potential) of the nucleons in the nucleus, to be compared with models of nuclear structure.
Various combinations of masses or binding energies can be made which highlight particular aspects of the nuclear structure.
For example, the \keyword{neutron separation energy}
\begin{equation}
S_n(Z,N) \equiv BE(Z,N) - BE(Z,N-1)
\end{equation}
is the minimum energy required to remove a neutron from a nucelus with $Z$ protons and $N$ neutrons.
A three-point mass difference is introduced in section~\ref{sec:CollectiveModes} in the context of pairing.

The mass of a nucleus can be measured in several ways, often involving the motion of an ion in the presence of a magnetic field.
Other techniques include time-of-flight measurements, or from the energy released in a reaction.
Experimental masses and related data are tabulated in, e.g.~\cite{Wang2021}.

\subsection{Densities and radii\label{sec:DensitiesRadii}}
The size and spatial distribution of a nucleus can be characterized by its charge density $\rho_{\rm ch}(\vec{r})$.
The size is measured by the root-mean-squared charge radius $\langle r_{\rm ch}^2\rangle^{1/2}$, where
\begin{equation}
    \langle r_{\rm ch}^2\rangle = \int d^3r \rho_{\rm ch}(\vec{r}) r^2.
\end{equation}
The charge density can be probed via elastic scattering of electrons.
More precisely, the elastic scattering differential cross section can be expressed in terms of a \keyword{form factor} $F(q)$, which depends on the momentum $q$ transferred from the electron to the nucleus:
\begin{equation}
    \frac{d\sigma}{d\Omega} = |F(q)|^2 \frac{d\sigma_{\rm pt}}{d\Omega},
\end{equation}
where $\frac{d\sigma_{\rm pt}}{d\Omega}$ is the elastic scattering cross section from a point charge.
In the first Born approximation, (i.e. leading order in perturbation theory), the form factor is the Fourier transform of the charge density.
However, if the sub-leading terms in the Born series are not negligible, the relationship between $F(q)$ and $\rho(\vec{r})$ is more complicated and model-dependent.
In particular, the density at small $r$ corresponds to large momentum transfer $q$, where cross sections are small and inelastic processes can become significant.
Consequently the density near $r\to 0$ is generally the most experimentally uncertain.

The charge radius may also be obtained from precise measurements of transition energies with muonic atoms.
For unstable isotopes, the charge radius can be inferred from the \keyword{isotope shift}, combined with a well-measured radius for the stable isotope of a given elemental chain.
Tabulated charge radii can be found in e.g.~\cite{Angeli2013}.

Neutrons are neutral, and so for the most part do not contribute to the charge density.
If one is interested in the matter density (i.e. protons and neutrons) within the nucleus, other probes are necessary.
Elastic scattering with a hadronic probe, such as nucleons, light nuclei, or pions, is sensitive to the distribution of protons and neutrons.
However, because the strong interaction is much stronger than the electromagnetic interaction, the first Born approximation is generally inadequate, and the extracted matter densities are much more model-dependent than charge densities~\cite{Tanihata1985,ZenihiroPRC2010}.

In principle, the neutron distribution can be probed by the weak interaction.
However, this is very challenging experimentally because the cross sections are small.
To date, weak interactions have been used to measure neutron radii in $^{48}$Ca and $^{208}$Pb~\cite{PREXII,CREX}, though still with significant uncertainty.
For more detail, see~\cite{GnechThis}.

\subsection{Excited states}
A direct consequence of the fact that the nucleus has structure is the possibility to excite a nucleus to a state of higher energy.
As in atoms and molecules, these excited states can de-excite by emitting a photon (in the nuclear case, the photons are called \keyword{gamma rays}, with energies of $\sim 1$~MeV).
By preparing a nucleus in an excited state and measuring the energies of the emitted gamma rays, it is possible to reconstruct the spectrum of low-lying excited states in a given nucleus.
These states are characterized by their excitation energy, as well as any conserved quantum numbers like angular momentum and parity (see section~\ref{sec:Symmetries}).
The determination of the angular momentum of a state can be achieved e.g. with laser spectroscopy\cite{Campbell2016} or by carefully analyzing the angular distribution of gamma rays~\cite{Smith2019}.
Guidelines for what evidence should be used for assigning angular momentum and parity may be found in~\cite{NDSpage}.

At higher energies (usually beyond a few MeV), the number of excited states often gets so high that it is no longer practical to tabulate individual states, and it becomes more meaningful to discuss a \keyword{density of states}.
Several types of nuclear reactions, such as neutron capture, often proceed through intermediate states where the level density is high; in these cases the density of states is an important input for predicting reaction rates~\cite{RichardThis}.

\subsection{Electromagnetic moments and transition rates}

Electromagnetic probes of nuclear structure are particularly useful because the electromagnetic interaction is well-understood, and because it is much weaker than the strong interaction between nucleons so that the nucleus is only mildly disturbed by the measurement.

Beyond the radial charge density $\rho_{\rm ch}(r)$, the charge may be distributed in the nucleus in a way that is not spherically symmetric.
As in classical electrodynamics~\cite{JacksonEandM}, it is convenient to decompose $\rho_{ch}(\vec{r})$ into multipole moments:
\begin{equation}
   \mathcal{M}_{E\lambda} \propto \int \!d^3r\, \rho_{\rm ch}(\vec{r}) r^\lambda Y_{\lambda \lambda}(\theta,\phi)
\end{equation}
where $Y_{\ell m}$ is a spherical harmonic.
Since the size of the nucleus will be small compared the length scale of most probes, the most important moments are those with low $\lambda$, due to the factor $r^\lambda$.
In addition, because the strong and electromagnetic forces conserve parity (see section~\ref{sec:Symmetries}), only even $\lambda$ can occur.
The $\lambda=0$ (monopole) moment simply corresponds to the total charge, so the most important electric multipole moment is thus $\lambda=2$ (quadrupole).

The electric quadrupole moment of a nucleus can be determined by measuring the interaction energy of a nucleus with an electric field gradient, typically generated by the charge distribution within a molecule.
Consequently, the extraction of the quadrupole moment often depends on a theoretical calculation of the electric field gradient, which can generally be done to sufficient precision for the purposes of nuclear structure.
Tabulated electric quadrupole moments can be found e.g. in ref.~\cite{Stone2013}.

The protons moving in the nucleus generate an electric current density which generates a magnetic field, and both protons and neutrons carry an intrinsic spin magnetic dipole moment.
Consequently, in addition to charge density, nuclei have a magnetization density, which can also be expanded into multipole moments~\cite{JacksonEandM,BlattWeisskopf}.
In this case, only odd $\lambda$ are allowed by parity, and by far the most important magnetic moment is $\lambda=1$ (dipole).
Magnetic dipole moments can be measured by nuclear magnetic resonance (NMR), among other techniques.
Tabulated values can be found in ref.~\cite{Stone2019}.

As mentioned previously, a nucleus can transition from an excited state to a lower-energy state by emitting a gamma ray.
These transitions can be accurately treated in perturbation theory, yielding Fermi's golden rule for the decay rate~\cite{Sakurai}
\begin{equation} \label{eq:FermiGoldenRule}
    \mathcal{W}_{f\to i} = \frac{2\pi}{\hbar}| \langle \psi_f| V | \psi_i\rangle |^2 \rho(E_f)
\end{equation}
where $V$ is the perturbing potential (here the radiation field), and $\rho(E_f)$ is the density of final states.
Because of the rotational symmetry of the system, it is advantageous to decompose the radiation field into corresponding to the angular momentum carried by the photon.
The cartesian wave vector $\vec{k}$ and polarization $\vec{\epsilon}$ are replaced by two kinds modes of multipolarity $\lambda$, labeled electric ($E\lambda)$ and magnetic ($M\lambda$)~\cite{JacksonEandM,BohrMottelson1,EisenbergGreinerExcitation}.
Electric photons of multipolarity $\lambda$ have a parity $\pi=(-1)^{\lambda}$, while magnetic photons have $\pi=(-1)^{\lambda+1}$.
The transition rate may then be written in terms of reduced transition probabilities~\cite{RingSchuck,WongIntro,ZelevinskyVolya}
\begin{equation}  \label{eq:BElambda}
    B(x\lambda; i\to f) \equiv
    \sum_{M_f} |\langle \psi_f | \mathcal{O}_{x \lambda} | \psi_i\rangle |^2 =
    \frac{1}{2J_i+1} \left| \langle \psi_f \| \mathcal{O}_{x\lambda} \| \psi_i \rangle \right|^2
\end{equation}
where $M_f$ indicates the projection of the angular momentum of the final nuclear state, and the double bars on the right side of~\eqref{eq:BElambda} indicate a reduced matrix element.
Angular momentum and parity conservation yield \keyword{selection rules} for gamma decay, which can aid in the determination of the $J^{\pi}$ of nuclear states.
As in atomic physics, typical photon wavelengths are significantly larger than the size of the nucleus.
The radial component of the multipole fields is given by a spherical Bessel function $j_\lambda(kr)$ which goes as $(kr)^\lambda$ for $kr\ll 1$, so generally the lowest $\lambda$ allowed by the selection rules is dominant.

Weisskopf suggested an estimate of the transition probabilities~\eqref{eq:BElambda}, assuming a single nucleon participated in the transition~\cite{BlattWeisskopf}.
Electromagnetic transition rates are often reported in \keyword{Weisskopf units} corresponding to the single-particle estimate.
For example, the Weisskopf estimate for an electric multipole ($E\lambda$) transition is
$B(E\lambda)_W = \frac{1.2^{2\lambda}}{4\pi} \left(\frac{3}{\lambda+3}\right)^2 A^{2\lambda/3} e^2\,{\rm fm}^{2\lambda}$.
If a measured transition is considerably larger than one Weisskopf unit, this is an indication that the transition is \keyword{collective}, with many nucleons contributing coherently.
If the transition is considerably smaller than one Weisskopf unit, it may be an indication that some approximate symmetry or other effect is hindering the transition.
Tabulated data on $E2$ transitions in even-even nuclei can be found in~\cite{PritychenkoADNDT2016}.

The above discussion also applies, with appropriate modifications, to \keyword{beta decay}, a weak interaction process in which a neutron inside a nucleus transforms into a proton, emitting an electron and antineutrino.
(Depending on the energetics the reverse process, in which a proton transforms into a neutron plus a positrion and a neutrino, could also be possible).
In the case of beta decay, the wavelength of the lepton wave function is even longer than for photons and the centrifugal barrier strongly suppresses decays in which the leptons carry away orbital angular momentum; so the dominant cases, called ``allowed decays'', are when the leptons have their spins aligned or anti-aligned (called Gamow-Teller and Fermi decays respectively).
Fermi decays primarily populate isobaric analogue states (see section~\ref{sec:Symmetries}), while Gamow-Teller decays have more variability.
For more detail, see~\cite{RavlicThis,MadurgaThis}.

\subsection{Direct reaction cross sections\label{sec:DirectReactions}}
If a nuclear reaction proceeds on a time scale which is fast compared with the time scale for motion of nucleons within the nucleus, it is called a \keyword{direct reaction}~\cite{Satchler,ThompsonNunes}.
In a direct reaction, only one or a few nucleons in the nucleus participate, while the others are passive spectators, providing a probe of single-particle aspects of nuclear structure.

One example of a direct reaction is electron-induced proton knockout, often indicated $(e,e'p)$, in which an incident electron strikes a bound proton, transferring enough energy to eject it from its nucleus (the prime indicates that the outgoing electron has different energy and momentum than the incident one).
To the extent that more complicated processes---e.g. final-state interactions of the proton with other nucleons---can be neglected or modeled, this reaction provides a probe of what the proton was doing inside the initial nucleus.

To study the behavior of neutrons, which are not efficiently knocked out by an electron beam, an alternative probe is needed.
A common choice is the reaction $(p,p'n)$; another is $(p,d)$, in which an incident proton picks up a neutron from the target nucleus, resulting in an outgoing deuteron, $^{2}$H.
One can also perform reactions such as $(d,p)$ in which a nucleon is added to the target nucleus.
Pair transfer reactions such as $(t,p)$, where $t$ indicates a triton ($^{3}$H), can be used to probe pairing effects described in section~\ref{sec:CollectiveModes}.

Assuming the reaction process is properly modeled, various reaction probes should reveal a consistent picture of the behavior of nucleons within the nucleus.
To some extent this has been achieved for $(e,e'p)$ and $(d,^{3}{\rm He})$ reactions~\cite{KramerNPA2001}, but puzzles remain for other probes~\cite{GaardeNPA1983,AumannPPNP2021}.

\section{\label{sec:NNforce}The force between nucleons}

One of the pillars of the study of nuclear structure is the effort to understand the extent to which observed properties of nuclei are consequences of specific features of the interaction between nucleons, and what are generic many-body phenomena.
This motivates a summary of the features of nuclear forces.
(The terms ``forces'', ``potentials'', and ``interactions'' are often used interchangeably).
We begin with a qualitative description, followed by a more microscopic picture in terms of meson exchange, and a discussion of three-body forces.
A more detailed discussion of nuclear forces can be found in~\cite{EkstromThis,ElsterThis,HebelerThis}.

\subsection{Qualitative features}
In the Standard Model of particle physics, the ``strong force'' refers to the interaction between quarks and gluons, as described by quantum chromodynamics (QCD).
Gluons couple to the color charge of quarks, in a manner analogous to photons coupling to the electric charge.
Nucleons are a color-neutral bound state of three quarks, in the same sense that a hydrogen atom is a charge-neutral bound state of an electron and a proton.
So to a leading approximation there is no interaction between well-separated nucleons, just as there is negligible interaction between two neutral atoms.
In fact, there is an interaction between neutral atoms: quantum fluctuations lead to instantaneous dipole moments in the atoms, resulting in a residual dipole-dipole type interaction called the van der Waals force~\cite{KawaiNat2016}.
Analogously, the force between nucleons can be thought of as a residual van der Waals-like interaction based on the underlying strong interaction between quarks and gluons.

\begin{figure}\centering
    \includegraphics[width=0.7\textwidth]{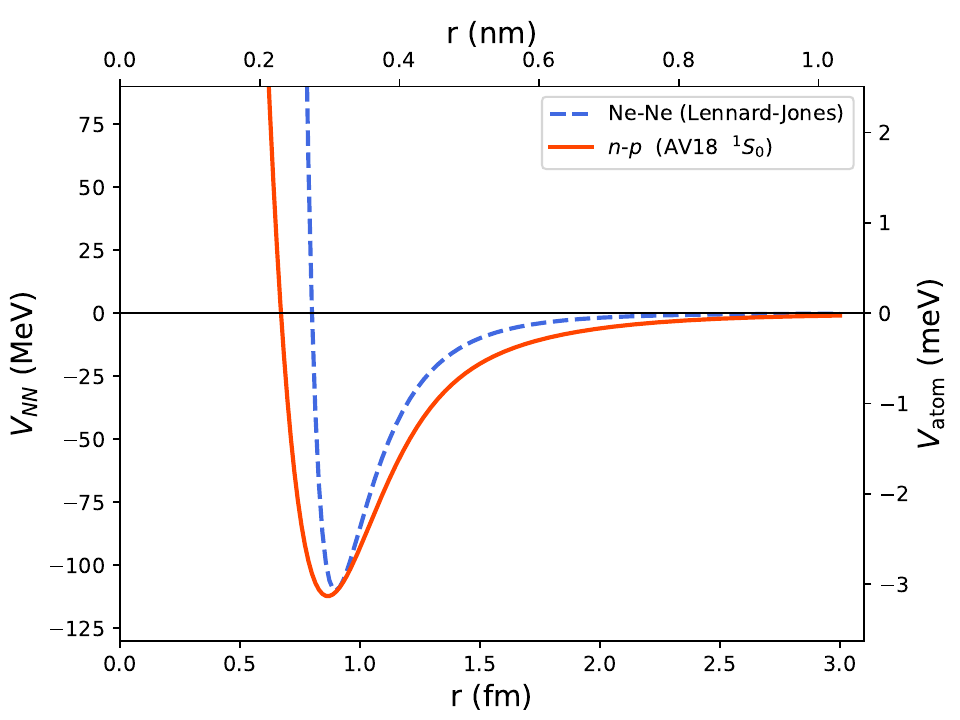}
    \caption{\label{fig:AV18vsLJ}Comparison of neutron-proton potential from the Argonne V18 model ($^1S_0$ channel)~\cite{WiringaPRC1995}, and the potential between two neon atoms modeled with the Lennard-Jones form~\cite{PchemBook}.}
\end{figure}

As shown in Fig.~\ref{fig:AV18vsLJ}, once adjusted to the relevant length and energy scales, the nuclear interaction and van der Waals interactions between neutral atoms look qualitatively similar.
Both fall off rapidly at large separation $r$, with attraction at intermediate distances, and strong repulsion as $r\to 0$.
For atoms, this repulsion can be understood as due to the Pauli principle once the electron clouds have significant overlap.
A similar interpretation can be made for nucleons: the quarks begin to overlap at separations comparable with the nuclear radius ($r_p\sim 0.7$~fm).

\subsection{Meson exchange\label{sec:MesonExchange}}
A major difficulty of the van der Waals picture of the nuclear force is that, for the low energies relevant for nuclear structure, quantum chromodynamics is highly nonperturbative, and quantitative predictions are very challenging to obtain. (See, for example,~\cite{NicholsonThis}).
At low energies quarks are confined in hadrons, and these hadrons become the relevant degrees of freedom.
As initiated by Yukawa~\cite{Yukawa1935}, the interaction between nucleons can be described as mediated by the exchange of mesons.

\begin{figure}
    \includegraphics[height=7em]{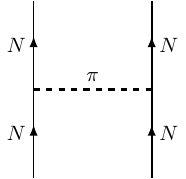}
    \qquad \quad
    \includegraphics[height=7em]{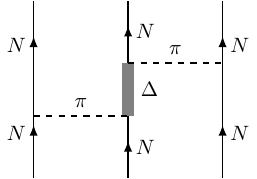}
    \qquad \quad
    \includegraphics[height=7em]{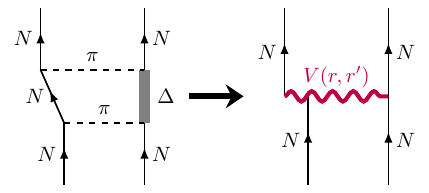}
    \caption{\label{fig:piondiagrams}(a) Diagram illustrating the one-pion exchange contribution to the $NN$ interaction. (b) The Fujita-Miyazawa diagram, contributing to a $3N$ interaction. (c) Elimination the $\Delta$ degree of freedom leading to a non-local effective interaction.}
\end{figure}

At lowest order (``tree level''), the exchange of a meson of mass $m$ with coupling $g$ to the nucleon yields a potential of the form $V(r)\sim g^2\frac{e^{-mr}}{r}$.
The range of the potential is inversely proportional to the meson mass, and so the longest-range part of the nucleon-nucleon potential is described by the exchange of the lightest meson: the pion, with $m_{\pi}\approx 140$~MeV, corresponding to a range of $\approx 1.4$~fm.
This process is illustrated by the Feynman diagram in Fig.~\ref{fig:piondiagrams}(a).
Heavier mesons contribute at shorter distances.
The next relevant mesons are the $\rho$ and $\omega$, with masses of 770 and 783 MeV, respectively, corresponding to a range of $\approx 0.25$~fm.

Sophisticated meson-exchange potentials are able to precisely reproduce NN scattering data~\cite{MachleidtCDBonn2001}.
Nevertheless, the meson-exchange picture also encounters difficulties in furnishing a fundamental theory of nuclear forces.
The issue is that at short distances an ever-increasing number of increasingly heavy (and short-lived) mesons will contribute with masses, coupling strengths, and form factors that are difficult or impossible to constrain independent of NN scattering.
Inevitably, meson-exchange potentials involve what amounts to a model of the short-range part of the potential.
It is difficult to assess whether this model corresponds to QCD, and it is unclear how to assess the uncertainty of theoretical predictions~\cite{Machleidt2011}.

Modern nuclear forces are formulated in the framework of effective field theory, in particular \keyword{chiral effective field theory} which incorporates the spontaneously-broken chiral symmetry of QCD~\cite{Machleidt2011,HammerRMP2020,Epelbaum2RMP009}.
In chiral effective field theory, the long-range part of the potential is due to pion exchange, while the short distance physics is systematically expanded in powers of $Q/\Lambda_\chi$, where $Q\sim m_\pi$ is the typical momentum scale in nuclear physics and $\Lambda_\chi\sim m_\rho$ is the scale of unresolved physics.
An essential feature of effective field theory is the possibility to estimate the size of omitted terms by using \keyword{naive dimensional analysis}~\cite{Manohar1984,Friar1997,HammerRMP2020}.
For more details, see~\cite{EkstromThis,ElsterThis,HebelerThis}.

\subsection{Composite structure of the nucleon}

From the point of view of QCD, nucleons are composite particles, made of quarks.
Consequently, nucleons in the presence of other hadrons can be deformed or polarized; equivalently, we may consider nucleons being excited to other particles.
The lowest such nucleon excitation is the $\Delta$ baryon, with mass $M_{\Delta} \approx 1232$~MeV.
When treating nucleons as fundamental degrees of freedom, we neglect the possibility that they are excited.
Such processes, when expressed in terms of purely nucleonic states, inevitably lead to \keyword{many-body forces}, which are forces betwen $A>2$ nucleons which cannot be decomposed into pairwise interactions.
An example, first suggested by Fujita and Miyazawa in 1957~\cite{FujitaMiyazawa1957}, is illustrated in Fig.~\ref{fig:piondiagrams}(b).
In this diagram two nucleons interact by pion exchange, which leads to the middle nucleon being excited to a $\Delta$ baryon.
The $\Delta$ baryon then decays back into a nucleon, emitting a pion which is absorbed by a third nucleon.
(A similar process can take place between just two nucleons as shown in Fig.~\ref{fig:piondiagrams}(c), a fact that is incorporated in chiral effective field theory).
Frequently, in the nuclear structure literature, one finds discussion of in-medium modification of the nuclear force, e.g.~\cite{Dickhoff1992}.
To the extent that the density dependence can be expressed as a Taylor expansion $V(\rho)=v_0+v_1\rho+v_2\rho^2+\ldots$, the leading correction to the {\it in vacuo} NN force is proportional to $\rho$ and corresponds to a 3N force~\cite{Holt2010,Hebeler2010,Lovato2011}.

Historically, 3N forces were omitted in nuclear structure calculations primarily due to the technical challenge of including them.
As many-body methods and available computational resources became more sophisticated, it became clear that 3N forces are a critical ingredient in the reproduction of the binding energies and spectra of light nuclei~\cite{Wiringa2002,NavratilPRL2007} and the saturation of nuclear matter (see section~\ref{sec:LiquidDrop}).

A second consequence of nucleon compositeness is that nuclear interactions will in general be \keyword{non-local}~\cite{Siemens1995,Machleidt1996}.
This means that rather than simply depending on the separation of two particles, represented as $V(r)$, the potential will connect initial and final states with different separations, i.e. $V(r,r')$.
Figure~\ref{fig:piondiagrams}(c) provides a physical picture illustrating why this should be the case.
As in the Fujita-Miyazawa diagram, two nucleons interact via the exchange of a pion, exciting one of the nucleons to a $\Delta$ baryon.
The $\Delta$ then emits a pion which is absorbed by the first nucleon; however, in the intervening time between the two pion exchanges, the nucleons have continued to move, so the absorption and emission occurs at different separation $r$.
Eliminating the intermediate $\Delta$ particle produces an instantaneous effective potential that is non-local.

\subsection{Insight from the Renormalization Group}
The typical momentum of nucleons inside a nucleus is given by the \keyword{fermi momentum} $k_{F}=(\frac{3\pi^2}{2}\rho)^{1/3}\sim 1.3~{\rm fm}^{-1}$, so nuclei cannot resolve features of the nuclear force at length scales below about 1~fm.
The short-distance part of the potential certainly has an impact, but there are infinitely many parameterizations of the short-distance piece that yield the same low-energy NN scattering.
This is the central idea behind the \keyword{renormalization group}~\cite{Wilson1983,Furnstahl2013}, and an essential element making effective field theory useful~\cite{Weinberg1983,Lepage1997} (see also~\cite{ElsterThis,EkstromThis,HebelerThis})

While many parameterizations of the NN potential yield the same results for the two-body problem, they will not be equivalent for $A\geq 3$.
In general, they will each require a different 3N force~\cite{Polyzou1990}.
As a practical matter, the equivalent NN potentials will generally \emph{not} be equally convenient for solving the many-body problem.
``Hard'' interactions---those with a strongly repulsive core---are generally more challenging to use than ``soft'' potentials without a repulsive core.
The similarity renormalization group~\cite{BognerPPNP2010} (see also~\cite{HeinzThis}) provides a way to transform a hard potential into a soft one without changing the predictions for NN scattering.
The price paid is that the softer potentials are non-local and come with induced many-body forces (with 3N force being most important).
While this is a useful computational tool, it also clarifies the point that certain aspects of the nuclear force cannot be uniquely specified by comparison with experiment.
Put another way, the nuclear potential is not an observable quantity.

\section{Nuclear structure phenomena\label{sec:Phenomena}}

\subsection{The liquid drop model, saturation, and nuclear matter\label{sec:LiquidDrop}}

\begin{figure}
    \includegraphics[height=0.35\textwidth]{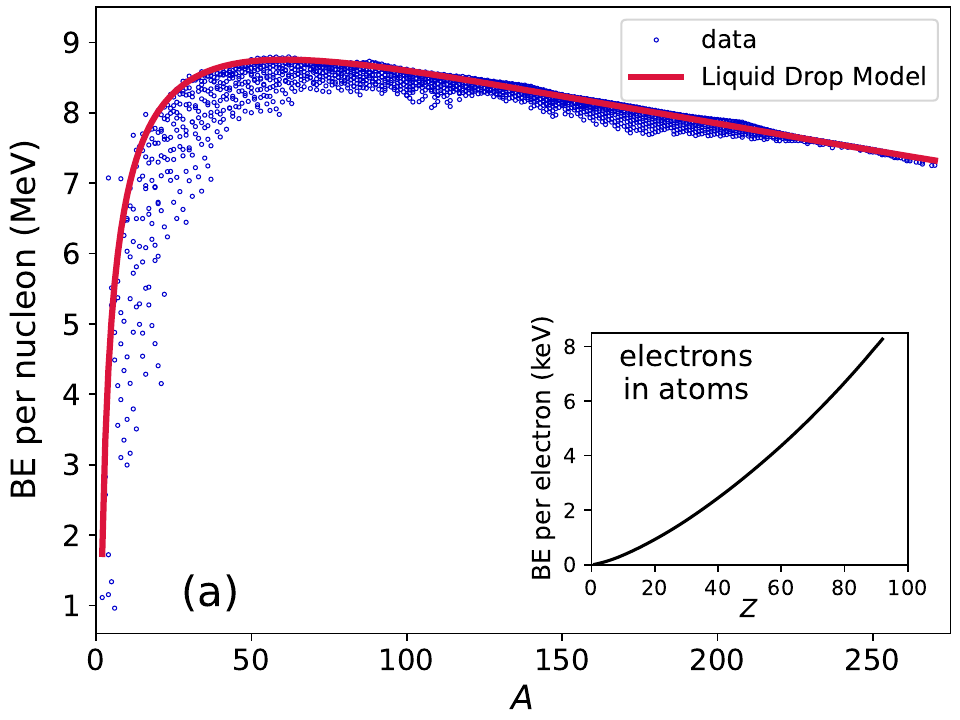}
    ~
    \includegraphics[height=0.35\textwidth]{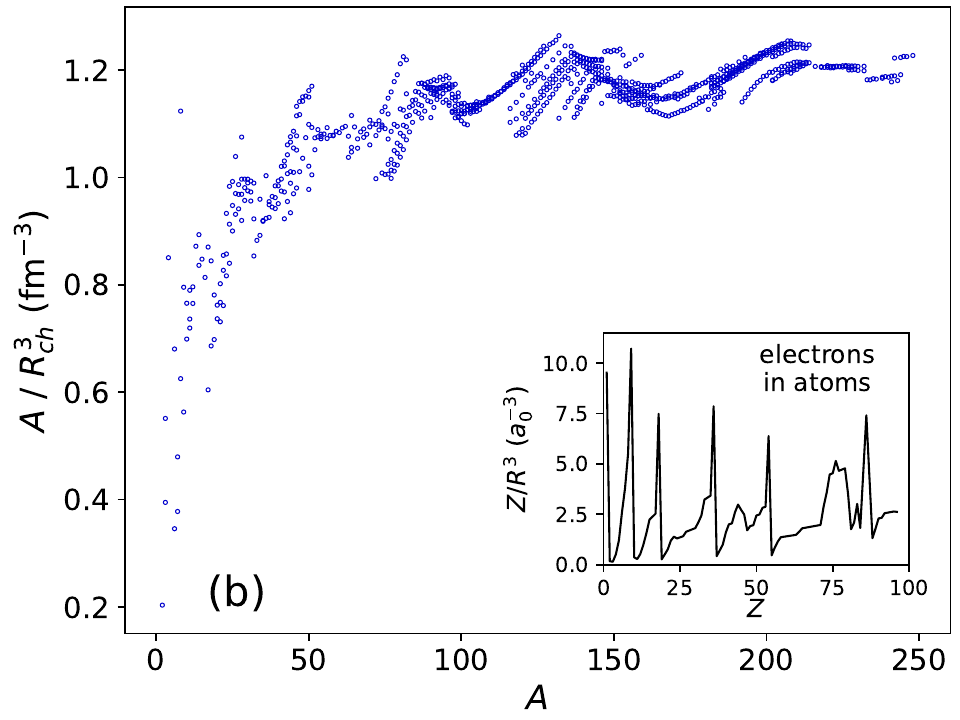}
    \caption{\label{fig:LDdata} (a) Experimental binding energy per nucleon, compared with the liquid drop model. (b) Number of nucleons divided by the charge radius cubed, which is roughly proportional to the density. The insets in both figures show the equivalent quantities for electrons in atoms, which do not manifest saturation.}
\end{figure}

The experimentally measured nuclear binding energies per nucleon $BE/A$ are shown in Fig.~\ref{fig:LDdata}(a).
For light nuclei, the average binding grows with increasing particle number until $A\sim 50$, at which point it saturates at approximately 8~MeV per nucleon.
The inset of Fi.~\ref{fig:LDdata}(a) shows the binding energy per electron in atoms, which grows with increasing atomic number without any sign of saturating.

A second signature of saturation can be observed in the systematics of the experimental root-mean-squared charge radius $R_{\rm ch}$, plotted in Fig.~\ref{fig:LDdata}(b) as the mass number $A$ divided by $R_{\rm ch}^3$, which should be proportional to the average density.
Beyond $A\sim 50$, $A/R_{\rm ch}^3$ is approximately constant, indicating a density that is independent of the number of nucleons.
This is again to be contrasted with the behavior of atoms, shown in the inset of Fig~\ref{fig:LDdata}(b), where the charge density varies by an order of magnitude with changing atomic number.

\begin{figure}
    \centering
    \includegraphics[width=0.95\linewidth]{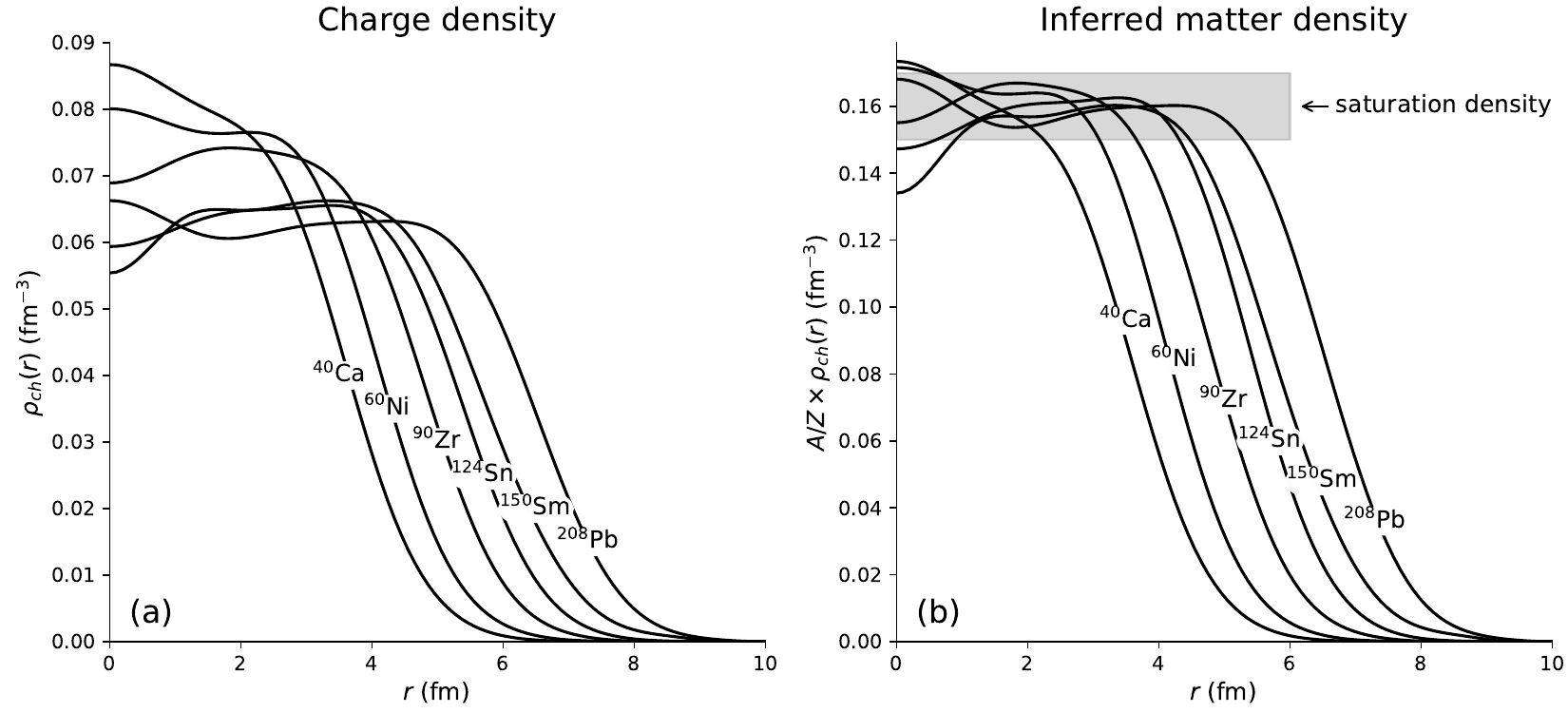}
    \caption{\label{fig:Densities}(a) Charge densities of selected nuclei, from~\cite{deVries1987}. (b) Inferred matter densities, obtained by scaling experimental charge densities by $A/Z$.}
    \label{fig:RhovsR}
\end{figure}

This picture is further supported by measurements of the charge density $\rho(r)$ from electron scattering, and the inferred matter density, shown in Fig.~\ref{fig:Densities}.
From $A=40$ to $A=208$, the central matter density does not significantly change; the nucleus grows larger, but not denser, just like a drop of incompressible liquid.

The binding energy of a nucleus can be modeled by the semi-empirical mass formula~\cite{WongIntro} (often called the \keyword{liquid-drop model})
\begin{equation} \label{eq:LiquidDrop}
\begin{aligned}
    BE(Z,N) = \,\alpha_V A - \alpha_S A^{2/3} - \alpha_A \frac{(N-Z)^2}{A} 
    &- \,\alpha_C \frac{Z^2}{A^{1/3}} + \Delta(Z,N).
    \end{aligned}
\end{equation}
Equation \eqref{eq:LiquidDrop} contains five terms.
The volume term proportional to $\alpha_V$ models the bulk binding of nuclear matter;
the surface term proportional to $\alpha_S$ models the fact that nucleons at the surface feel less binding than those in the bulk;
the asymmetry term models the effects of the Pauli exclusion principle, as well as the fact that the nuclear interaction between protons and neutrons is more attractive than the interaction between like nucleons;
the Coulomb term proportional to $\alpha_C$ models the long-range electrostatic repulsion between protons;
and the pairing tern $\Delta(Z,N)$, which can be parameterized in several ways, reflects the observation that nuclei with an even number of protons or neutrons are more bound.
Typical values, in MeV, are $\alpha_V=15.5$, $\alpha_S=17$, $\alpha_A=23$, $\alpha_C=0.7$.
(In this section we omit the pairing term to simplify the discussion).

\begin{figure}
    \centering
    \includegraphics[width=0.45\linewidth]{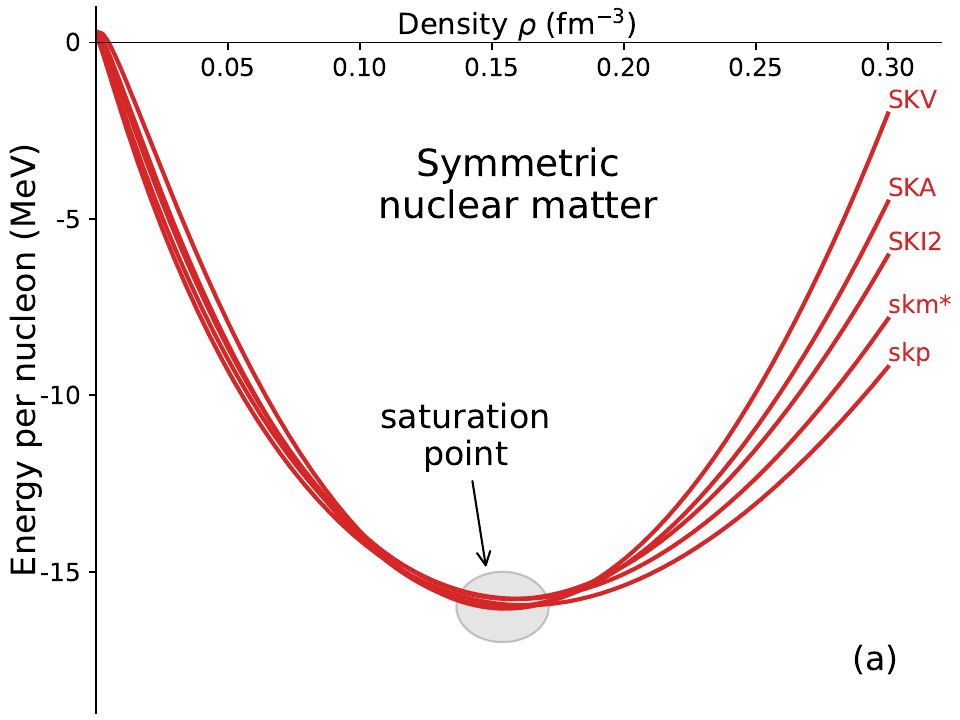}
    ~~
    \includegraphics[width=0.45\linewidth]{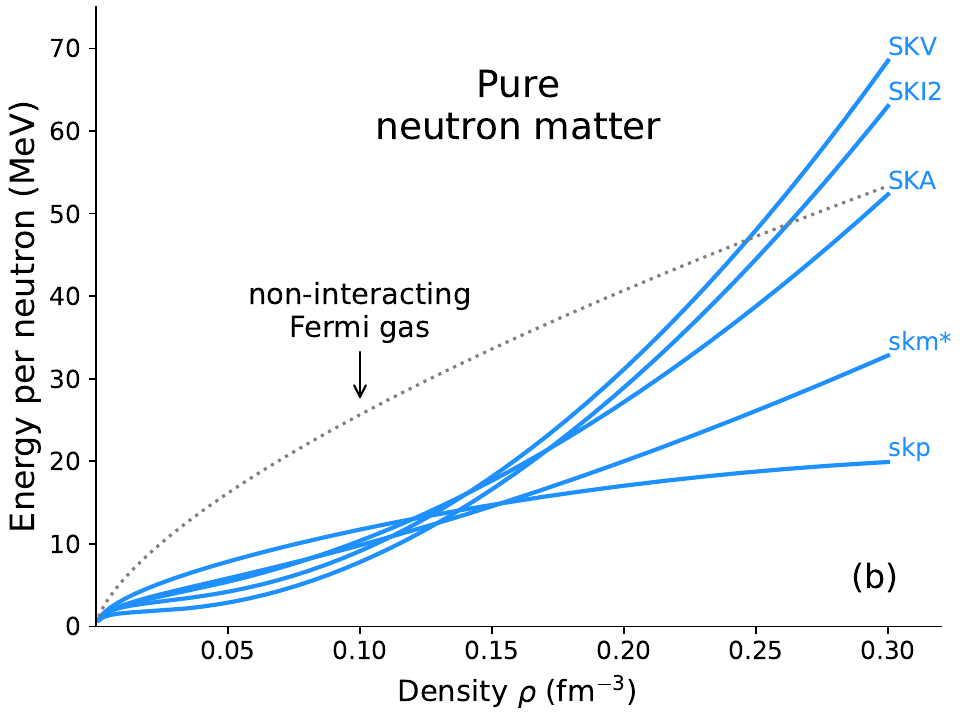}
    ~
    \caption{\label{fig:EOS}(a) Equation of state for symmetric nuclear matter computed with several Skyrme energy density functionals~\cite{SKV,SKA,SKI2,SKM,SKP}. (b) Equation of state of pure neutron matter with the same Skyrme functionals. The dotted line indicates the behavior of a non-interacting Fermi gas, $E/A\propto \rho^{2/3}$.}
    
\end{figure}

If we imagine the limit $A\to \infty$ and ignore the Coulomb term, then we obtain a system called \keyword{infinite nuclear matter}.
The liquid drop formula~\eqref{eq:LiquidDrop} predicts that in this limit the binding energy per nucleon approaches $\alpha_V\approx 16$~MeV, with all other terms vanishing.
The densities of finite nuclei shown in Fig.~\ref{fig:RhovsR} imply that the energy should be minimized at the \keyword{saturation density} $\rho_0\approx 0.16~{\rm fm}^{-3}$.
It is a non-trivial requirement that a model of nuclear forces should reproduce the empirical saturation point in calculations infinite nuclear matter.
The full curve of energy as a function of density defines the \keyword{nuclear matter equation state} (one can also consider dependence on e.g. temperature or proton fraction).
The equation of state obtained with several mean-field models is shown in Fig.~\ref{fig:EOS}(a).
The behavior is qualitatively consistent with the behavior of the nuclear force in Fig.~\ref{fig:AV18vsLJ}; at low densities, nucleons mostly experience the attractive long-range component of the force, while at high densities the repulsive short-distance component dominates.
Quantitatively, modern calculations find that 3N forces are also essential to obtain the repulsion at high densities needed for saturation~\cite{Day1983,Carlson1983,Hebeler2011}.

We may also consider the limit $Z=0$, $N=A\to \infty$, called \keyword{pure neutron matter}, which is of central importance for understanding the physics of neutron stars~\cite{DexheimerThis,PageThis}.
In this case, the liquid drop model binding energy approaches $BE/N\approx \alpha_V-\alpha_A$.
From fits to data one generally finds that $\alpha_A>\alpha_V$, so we conclude that pure neutron matter is not bound at nuclear densities.
The curves shown in Fig.~\ref{fig:EOS}(b), obtained with mean-field models fit to finite nuclei, all predict unbound neutron matter.
Away from the saturation point the model dependence is large.
Calculations of neutron matter based on microscopic interactions yield a more tightly constrained equation of state up to the saturation density~\cite{WiringaPRC1988,HebelerApJ2013,DrischlerARNPS2021,HuNP2022}. 
This can be understood based on the fact that the leading short-range 3N forces, which are the largest source of uncertainty in nuclear matter, do not contribute due to the Pauli principle; consequently neutron matter is well constrained by NN scattering data.
For a recent review of calculations of nuclear matter and applications to neutron stars, see ~\cite{DrischlerARNPS2021,LattimerARNPS2021,DexheimerThis}.

\subsection{Shell structure\label{sec:ShellStructure}}
\begin{figure}
    \centering
     \includegraphics[width=0.52\linewidth]
     {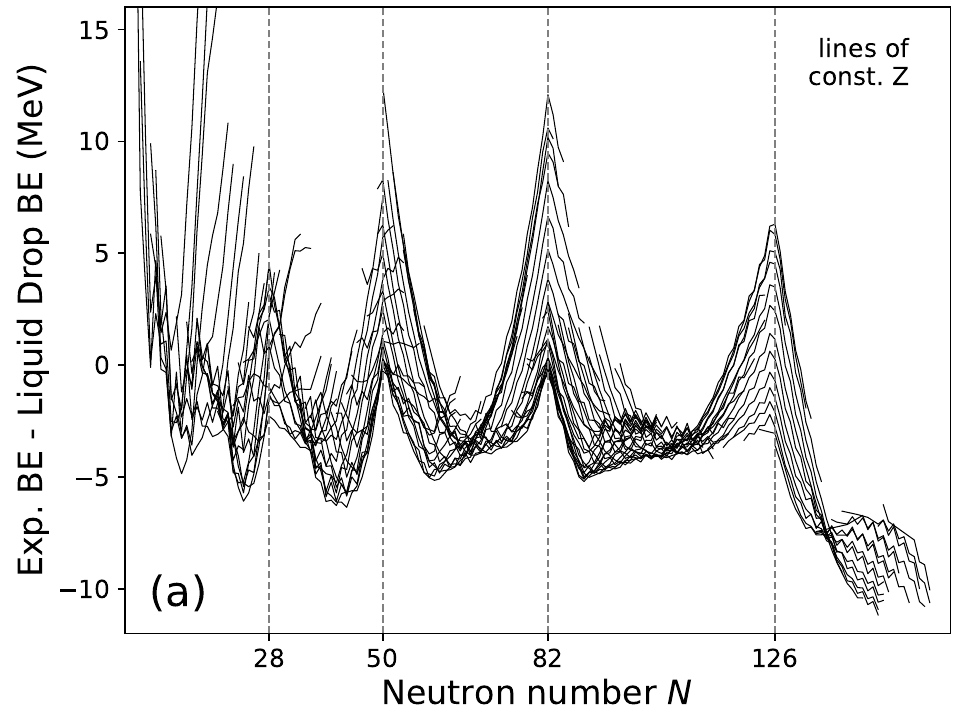}\quad 
    \includegraphics[width=0.40\linewidth]{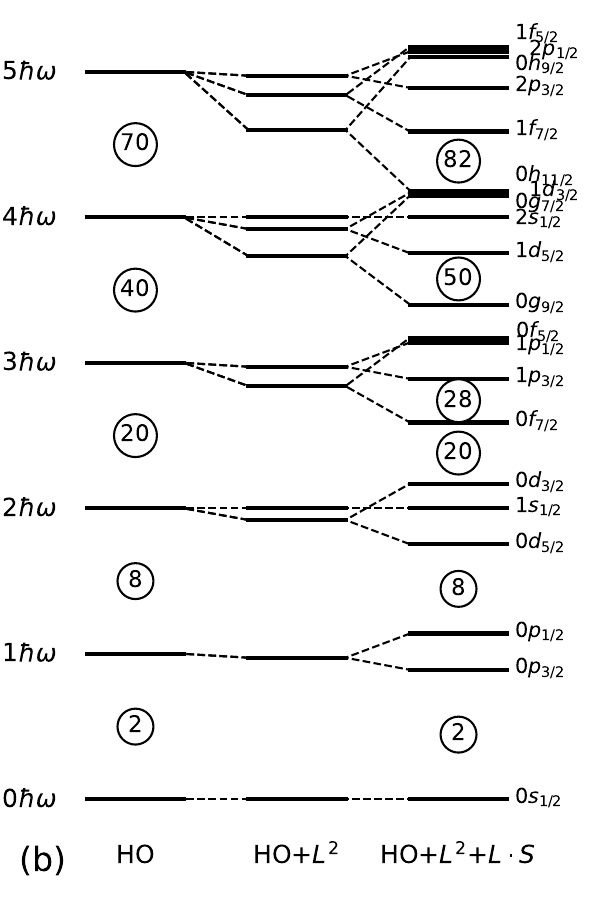}
    \caption{(a) residual of experimental binding energies from the liquid drop model fit. (b) Shell gaps obtained for a 3D harmonic oscillator potential, with the addition of terms proportional to $L^2$ and $L\cdot S$.}
    \label{fig:magicnumbers}
\end{figure}

While the notion of a nucleus as a classical liquid drop is a useful starting point, quantum mechanics is crucial for understanding many features.
Probably the most important quantum effect in nuclei is \keyword{shell structure}.
If the liquid drop prediction~\eqref{eq:LiquidDrop} is subtracted from the experimental binding energy data, the residual shown in Fig.~\ref{fig:magicnumbers}(a) demonstrates clear peaks at specific neutron numbers (peaks are also found at the corresponding proton numbers).
The existence of these so called \keyword{magic numbers} $N=2,8,20,28,50,82,126$ can be understood by assuming that each nucleon moves in a mean field potential generated by all the other nucleons in the nucleus.
Solving the one-particle Schr\"odinger equation yields a discrete set of bound states.
Nucleons obey the Pauli exclusion principle, so the lowest $Z$ proton and $N$ neutron orbits will be occupied.
The spikes in Fig.~\ref{fig:magicnumbers}(a) correspond to larger-than-average gaps in the one-particle energy spectrum.

Obtaining gaps at the specific numbers observed in nuclei requires a model of the mean-field potential.
Since the nucleon-nucleon interaction is short-ranged, the mean-field potential should roughly follow the density distribution in Fig.~\ref{fig:RhovsR}.
For well-bound states, this can be approximated by a 3D harmonic potential $V=\tfrac{1}{2}m\omega^2 r^2$.
The eigenstates of this potential are labeled by orbital angular momentum $l$, its projection $m_l$, and the radial quantum number $n=0,1,2,\ldots$ which counts the number of radial nodes\footnote{A common alternative convention is for $n$ to start at 1.}.
In addition, there are two spin states $m_s = \pm \tfrac{1}{2}$.
The energies are $E_{\mathcal{N}}=(\mathcal{N}+\tfrac{3}{2})\hbar\omega$, with $\mathcal{N}\equiv 2n+l$,
and the degeneracies of these levels are $D_\mathcal{N}=2(\mathcal{N}+1)(\mathcal{N}+2)$, where the factor 2 comes from the spin projections.
Thus we predict large gaps at 2, 8, 20, 40, 70\ldots, which reproduces the first three magic numbers but fails for the remaining ones.
The density distribution is better approximated by a rounded square well, which can be mocked up by adding a potential proportional to $L^2$, but this does not change the predicted magic numbers.

The critical ingredient for the magic numbers beyond $20$ is a spin-orbit potential $V_{\rm so}\propto \vec{L}\cdot \vec{S}$~\cite{MayerJensen1955}.
In the presence of a spin-orbit potential, the projections $m_l$ and $m_s$ are no longer conserved, and it is advantageous to use orbits in which the nucleon spin and orbital angular momentum are coupled to a total angular momentum $j=l\pm\tfrac{1}{2}$ with projection $m_j=m_l+m_s$.
It is common to label the orbits with spectroscopic notation $n l_{j}$.
The orbital angular momentum $l$ is denoted by $s,p,d,f\ldots$ for $l=0,1,2,3\ldots$ so that an orbit with $n\!=\!0,\, l\!=\!3,\, j\!=\! \tfrac{7}{2}$ would be labeled $0f_{7/2}$.
A given $nl_j$ orbit has degeneracy $2j+1$ corresponding to the possible values of $m_j$, which do not affect the energy.
The parity of the orbit can be obtained from its orbital angular momentum $\pi=(-1)^l$.
The magic number $N=28$ can be understood as due to the spin orbit splitting between the $0f_{7/2}$ and $0f_{5/2}$ levels, as can be seen in Fig.~\ref{fig:magicnumbers}(b).

Initially, it might seem counterintuitive that nucleons would follow regular orbits, given the strength of the nucleon-nucleon force and the high density of the nucleus.
However, this too can be understood as a consequence of the Pauli exclusion principle.
In order for two deeply-bound nucleons to scatter off one another, they should have available final states to scatter into; but the low-lying final states are already occupied, so the scattering is suppressed.
This point is discussed further below and in section~\ref{sec:SRC}.

\noindent
{\bf Experimental signatures of magic numbers}

In addition to explaining Fig.~\ref{fig:magicnumbers}(a), the shell model picture predicts the ground state angular momentum, parity, and magnetic moments of odd-mass nuclei, as well as the systematics of long-lived excited states (called isomers, in analogy with molecules) ~\cite{MayerJensen1955}.
We assume that like nucleons combine to make $J=0$ pairs (see section~\ref{sec:CollectiveModes}) so that the angular momentum and parity is determined by the last unpaired nucleon.
For example, we may consider the nucleus $^{41}$Ca as a single neutron on top of the doubly-magic (N=Z=20) $^{40}$Ca.
According to the shell model, the lowest-energy available state for the additional neutron is the $0f_{7/2}$ orbit, so the ground state of $^{41}$Ca should have $J^{\pi}=\tfrac{7}{2}^-$, which is indeed the case experimentally.
A more thorough discussion of the experimental evidence for nuclear shell structure can be found in~\cite{MayerJensen1955}.

{\bf Correlations and the residual interaction}

Above, it was argued that the Pauli principle should suppress scattering of nucleons out of their mean-field orbits.
However, this argument does not hold for nucleons near the fermi surface (i.e. the occupied states with the highest energy), since there are nearby unoccupied states available.
In this case, the residual interaction (i.e. the component that cannot be expressed as a mean field) can play an important role for nuclear structure.

When two nucleons, initially in orbits $\phi_1$ and $\phi_2$, can interact and scatter into two different orbits $\phi_3,\phi_4$, then in general the probability to find a nucleon in orbit $\phi_3$ will depend on whether or not there is a nucleon in orbit $\phi_4$.
This means the behavior of the two nucleons is \keyword{correlated} or, equivalently, that the nucleons are \keyword{entangled}.
This may also be stated without reference to orbits.
Using $x$ to label position and spin/isospin degrees of freedom, the probability to find a nucleon at $x_1$ is $P(x_1)$.
The joint probability to simultaneously find a nucleon at $x_1$ and $x_2$ is $P(x_1,x_2)$.
If $P(x_1,x_2)\neq P(x_1)P(x_2)$, then the wave function is correlated (or entangled).
Note that even in a simple shell model configuration, the Pauli principle leads to correlations such that two identical fermions can't be found at the same point.
Typically, when one speaks of correlations in a wave function, one means correlations beyond those implied by the Pauli principle.

One way to express correlated wave functions is as a superposition of different configurations of $A$ nucleons in a set of single-particle orbits.
If we label one such $A$-body configuration $\Phi_i$, then the fully correlated wave function $\Psi$ can be written
\begin{equation} \label{eq:CI}
    \Psi = \sum_i c_i \Phi_i
\end{equation}
where the $c_i$ are amplitudes determined by solving the $A$-body Schr\"odinger equation, including the residual interaction.
If more than one configuration $\Phi_i$ makes a significant contribution to $\Psi$, then we say the wave function is correlated. 
This ``configuration interaction'' approach is a powerful, though expensive, way to treat the many-body problem (see \cite{McCoyThis}).
But it also provides a framework for approximate methods~\cite{HergertThis,FossezThis,HeinzThis,BaccaThis}, as well as a conceptual picture for understanding correlations~\cite{Suhonen,Talmi1993,CaurierRMP2005}.

{\bf Shell evolution}

As mentioned above, the specific values of the magic numbers depend on the form of the mean-field potential.
It turns out that the shell gaps are not universal for all nuclei.
Due to the short-range nature of the nuclear interaction, the mean field potential should follow the density $\rho(\vec{r})$.
If the density assumes a non-spherical shape, then so will the mean field.
This breaks the degeneracy of the different $m_j$ projections in a given $nl_j$ orbit, and can lead to new shell gaps.
This is described below in section~\ref{sec:CollectiveModes}.

The shell gaps can also change with the relative number of protons and neutrons~\cite{OtsukaRMP2020}.
The essential reason is that while the proton-proton and neutron-neutron interactions are essentially the same due to isospin symmetry, the proton-neutron interaction can act in channels (isospin $T=0$) that would be forbidden by the Pauli principle in the like-particle case.
In terms of the meson-exchange picture described in section~\ref{sec:MesonExchange}, two protons (or two neutrons) can only exchange a neutral pion, while a proton-neutron pair can also exchange charged pions.
Consequently, a proton will feel a different mean field in, e.g., $^{40}$Mg ($Z=12$, $N=28$) compared with $^{40}$Ca ($Z=N=20$).

More specifically, due to the spin-isospin structure of the one-pion exchange, the occupied proton orbits impact the spin-orbit splittings of neutron orbits, and vice versa~\cite{OtsukaPRL2005}.
The spin-orbit potential was essential for establishing all of the magic numbers beyond 20, and changing the number of protons can produce an important shift in the neutron shell structure.
For this reason, naively extrapolating phenomenological models fit near stability to nuclei with extreme proton-neutron ratios (relevant e.g. for nuclear astrophysics) can lead to significant error.

Three-nucleon forces play an important role in the evolution of shell structure.
Several investigations have found that 3N forces contribute significantly to the mean-field spin-orbit potential~\cite{FujitaMiyazawa1957b,AndoPTP1981,FukuiPLB2024,DingPRL2026}.
In addition, 3N forces encode effects of the Pauli exclusion principle which can affect shell structure and e.g. determine which isotopes of a given element are bound~\cite{OtsukaPRL2010}.

Finally, weak-binding effects (see section~\ref{sec:WeakBinding}) can also modify shell structure.
As the binding energy of a single particle orbit approaches zero, its wave function becomes extended is space.
Due to the centrifugal barrier, orbits with high $l$ ``see'' less of the attractive nuclear potential.
The net effect is that near threshold low-$l$ orbits drop down in energy relative to higher-$l$.
A classic example of this is $^{11}$Be ($Z=4,N=7$), with a neutron separation energy of only 500~keV, where the ground state has $J^\pi=\tfrac{1}{2}^+$ (corresponding to $l=0$), rather than the $\tfrac{1}{2}^-$ predicted by the shell model picture of Fig.~\ref{fig:magicnumbers}(b).

\subsection{Collective Modes: Pairing, Vibrations, and Deformation\label{sec:CollectiveModes}}

While the configuration-interaction framework~\eqref{eq:CI} is very general, it can yield very complicated descriptions of nuclear wave functions.
In many cases, a simpler description can be obtained by considering the collective behavior of many nucleons.
The most important collective modes are pairing, surface vibrations, and static deformation.
More detailed discussions can be found in~\cite{CoelloThis,AprahamianThis} and in textbooks~\cite{BohrMottelson2,RingSchuck,Rowe2010}.

\noindent
{\bf Pairing}

\begin{figure}
    \centering
    \includegraphics[width=0.5\linewidth]{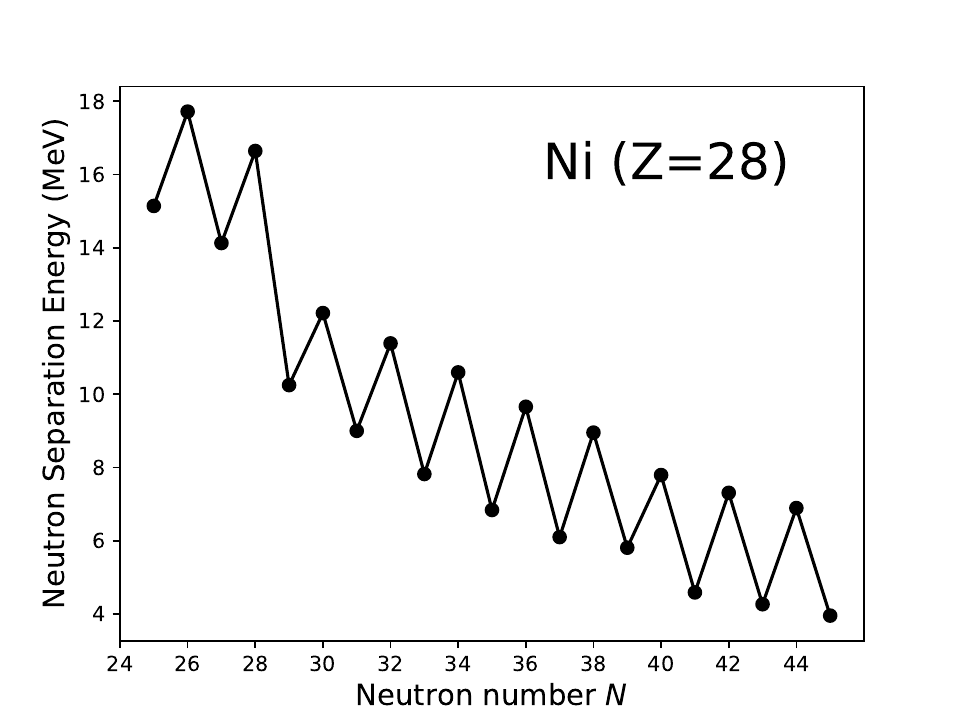}
    \caption{Neutron separation energy for nickel isotopes, displaying a clear odd-even staggering.}
    \label{fig:SnOddEven}
\end{figure}

All known nuclei with even $N$ and even $Z$, without exception, have a ground state with $J^\pi=0^+$.
Odd-even and odd-odd nuclei exhibit no such universality.
In the shell model, the assumption that the $J^\pi$ of odd-mass nuclei is determined by the last unpaired nucleon yields agreement with a vast amount of experimental data.
In addition, several observables, such as the one-neutron separation energy $S_n$ shown in Fig.~\ref{fig:SnOddEven} for the nickel isotopic chain, display a marked odd-even staggering.
Evidently, an even number of neutrons is more energetically favorable than an odd number, and likewise for protons.
This effect is often included in the liquid drop formula~\eqref{eq:LiquidDrop}, even though no such effect is predicted for a classical liquid drop.
The microscopic origin of these observations is related to the tendency of like nucleons to form \keyword{Cooper pairs} (typically with angular momentum $J=0$), in analogy with the electron Cooper pairs in conventional superconductors~\cite{BohrMottelsonPines}.
This tendency is due to the short-range attractive nature of the nuclear interaction.
Cooper pairs approximately behave like bosons, which means that multiple pairs can condense into the same quantum state, forming a superfluid~\cite{BohrMottelsonPines,BrinkBroglia,BrogliaZelevinsky2013, Rowe2010}.

A key observable related to pairing is the \keyword{empirical pairing gap}
, or three-point mass difference~\cite{BohrMottelson1,Suhonen}
\begin{equation}\label{eq:Delta3}
    \Delta^{(3)}_{n}(N) \equiv  \tfrac{(-1)^{N+1}}{2}\left[ BE(N+1,Z) +BE(N-1,Z) - 2BE(N,Z) \right] .
\end{equation}
The pairing gap quantifies how the binding energy of a given isotope with $N$ neutrons differs from the average of its two neighbors with $N\pm 1$ neutrons.
An analogous definition can be made for the proton gap $\Delta^{(3)}_p(Z)$ in terms of proton separation energies.
In terms of the pair condensate, $2\Delta$ corresponds to the energy required to break a pair.

\begin{figure}
    \centering
    \includegraphics[width=0.95\linewidth]{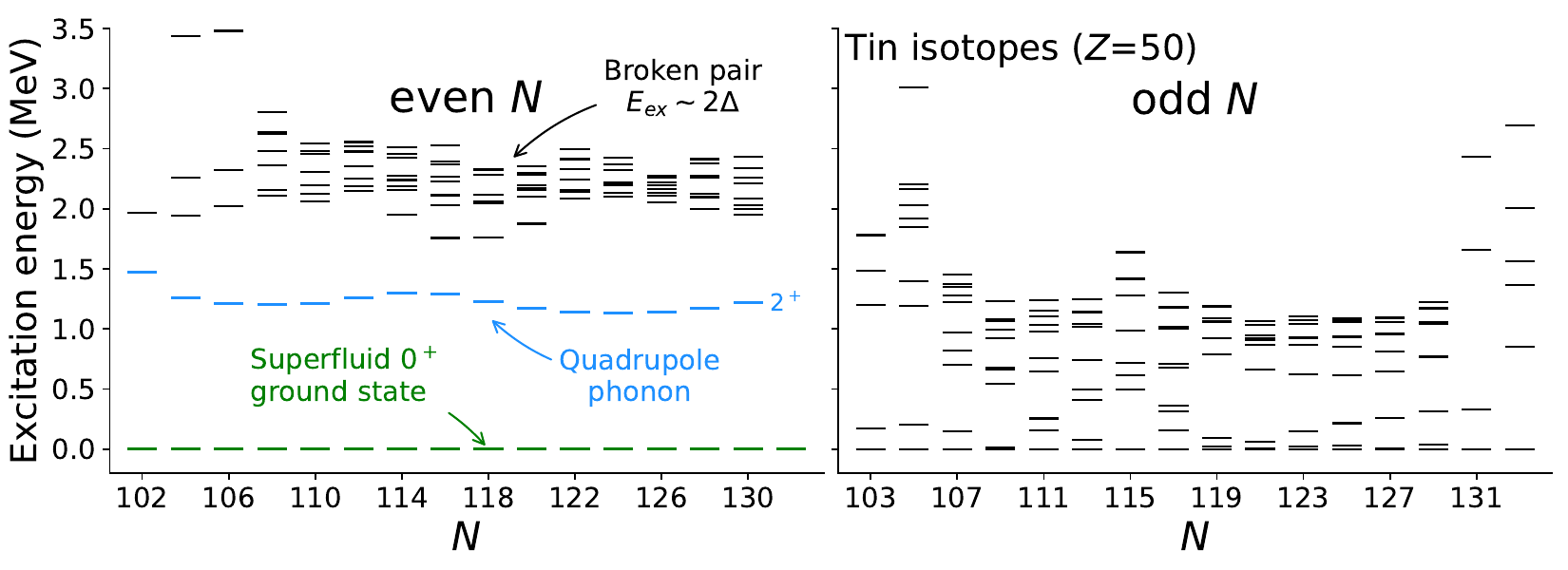}
    \caption{Excitation spectrum of isotopes of tin, from $^{102}$Sn to $^{132}$Sn with (a) even-$N$ and (b) odd-$N$.}
    \label{fig:SnSpectrum}
\end{figure}

For a given even-$N$ isotope, the lowest excitation energy should then be approximately $2\Delta$.
Fig.~\ref{fig:SnSpectrum} shows the excited-state spectrum for the tin isotopes, divided into even and odd $N$.
We find that for even $N$, the lowest excited states (excluding the first $2^+$ state) begin at an energy of about 2~MeV, which is not too far from the empirical $2\Delta\approx 2.5$~MeV obtained from the odd-even staggering of binding energies.
In stark contrast, excited states in the odd-$N$ isotopes begin around 100~keV, demonstrating the absence of an energy gap.

The lowest $2^+$ state in the even-$N$ isotopes can be interpreted as a collective quadrupole phonon (see discussion below).
This highlights the possibility of different collective modes to coexist in the low-lying spectrum.

Pairing effects in nuclei exhibit important differences from the superconductivity found in bulk electronic materials.
An essential feature is that nuclei are of finite size and have a finite number of particles.
The \keyword{coherence length} $\xi=\hbar v_F/2\Delta$, where $v_F$ is the fermi velocity and $\Delta$ is the pairing gap, measures the size of a Cooper pair.
In a metal, a typical size might be $\xi\sim 1 \mu\text{m}$, which is four orders of magnitude larger than the distance between atoms.
In a nucleus where $v_F\sim 0.3c$ and $\Delta\sim 1$~MeV, the coherence length is $\xi\sim 30$~MeV, which is much larger than the nuclear radius $R\sim 5$~fm, so that the size of a Cooper pair is given by $R$, rather than $\xi$~\cite{BrinkBroglia}.
The finite size of the nucleus also yields shell structure, as described in section~\ref{sec:ShellStructure}, so that the number of nucleons near the Fermi surface available to form Cooper pairs will depend on the number of nucleons relative to the nearest closed shell.

The approximate wave function proposed by Bardeen, Cooper, and Schriefer (BCS)~\cite{Bardeen1957} is a coherent state of Cooper pairs, which does not have a well-defined particle number $N$.
For macroscopic systems where $N\to\infty$, the resulting particle-number fluctuations are generally a negligible effect.
However, in nuclei the particle-number fluctuations are significant and it is often necessary to project back onto good particle number~\cite{RingSchuck}.
The finite particle number tends to somewhat wash out sharp phase transitions observed in macroscopic systems, so that experimental signatures of superfluidity are more subtle~\cite{BrinkBroglia}.

{\bf Vibrations}

Within the liquid drop picture, one may consider the possibility of a drop with a non-spherical shape.
Since a sphere minimizes the surface-to-volume ratio for a given density, a non-spherical shape should increase the repulsive contribution of the surface term in the liquid drop formula \eqref{eq:LiquidDrop}.
Classically, this would mean that the ground state of the nucleus is spherical.
Quantum mechanically, the ground state will in general be a superposition of different shapes.
More formally, the surface of the drop can be parameterized as a distance $R(\theta,\phi)$ depending on the angles $\theta,\phi$, and this can be expanded in terms of spherical harmonics
\begin{equation}
    R(\theta,\phi) = R_0\left[ 1+\sum_{\lambda\mu} \alpha_{\lambda \mu}Y_{\lambda\mu}(\theta\phi) \right].
\end{equation}
A spherical shape would have all amplitudes $\alpha_{\lambda\mu}=0$.
The dipole mode $\lambda=1$ corresponds to a translation of the nucleus, and so is not relevant for the structure of the nucleus.
The lowest non-trivial multipole is then the quadrupole $\lambda=2$.

The amplitudes $\alpha_{\lambda\mu}$ can be regarded as dynamical variables.
Expanding the energy about the spherical minimum, the leading correction is proportional to $\alpha_{\lambda\mu}^2$, yielding a harmonic potential for small deformations.
Classically, a harmonic time dependence $\alpha_{\lambda \mu}(t)$ would describe a vibrating liquid drop.
Quantum mechanically, we can carry over the textbook solution to the 1D harmonic oscillator~\cite{GriffithsQM}, finding that the ground state wave function is a Gaussian in $\alpha_{\lambda \mu}$ centered on $\alpha_{\lambda\mu}=0.$
The excitation spectrum of the oscillator is evenly spaced in units of the energy quantum $\hbar\omega$: $E_n=E_0+n\hbar\omega$.
In the case of the liquid drop, the interpretation is that the vibrations have been quantized, and $n$ counts the number of \keyword{phonons}, with each phonon carrying energy $\hbar\omega$.
This phenomenon is analogous to the vibrational spectra in molecules.

In the shell model picture, a phonon can be understood as a coherent superposition of particle-hole excitations of out of the ground state~\cite{BohrMottelson2,RingSchuck}.
From an effective field theory point of view, the phonon is an emergent low-energy degree of freedom; one may perform an expansion in the ratio $Q/\Lambda$ where $Q$ is the phonon energy and $\Lambda$ is the breakdown scale given by the energy of a single particle-hole excitation~\cite{CoelloPRC2015}.
The fact that typically $\Lambda\sim 2Q$ means that it is uncommon to find unambiguous three-phonon states in nuclear spectra.

At higher energies, large-amplitude oscillations are observed, called \keyword{giant resonances}~\cite{BaccaThis}.
A classic example is the giant dipole resonance, in which protons and neutrons oscillate with opposite phase.
These resonances can be used to probe bulk properties of nuclear matter, for example the incompressibility $\kappa$ which characterizes the curvature of the equation of state at the saturation point in Fig.~\ref{fig:EOS}~\cite{Garg2018}.

{\bf Deformation}

\begin{figure}
    \centering
    \includegraphics[width=0.7\linewidth]{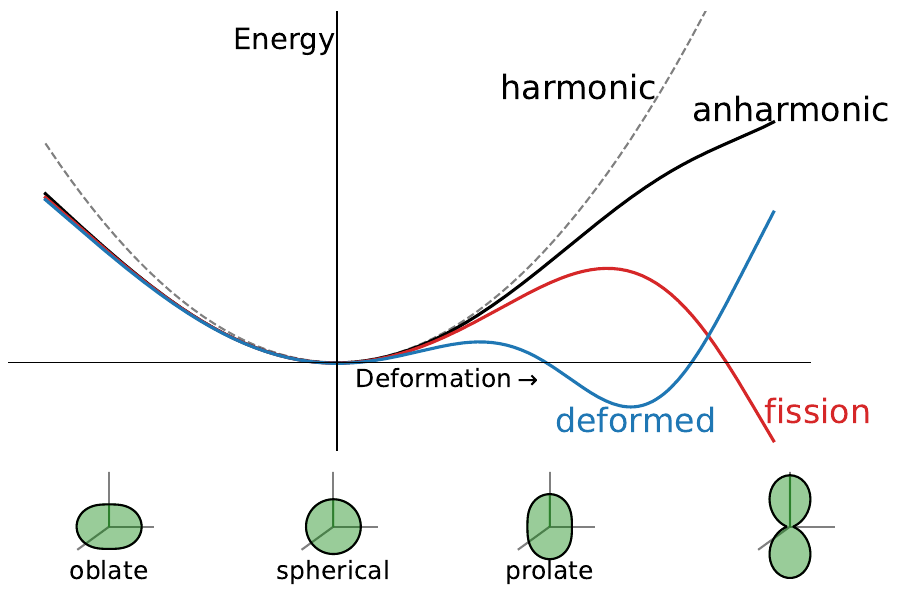}
    \caption{Schematic ``potential energy surface'' showing the energy of a liquid drop as a function of its quadrupole deformation.}
    \label{fig:PES}
\end{figure}

At larger amplitudes $\alpha_{\lambda\mu}$, the harmonic approximation will eventually break down.
For very large deformation, the liquid drop will split into two spherical drops resulting in nuclear \keyword{fission}.
This will be energetically favorable if the reduction in Coulomb repulsion outweighs the increased surface energy.
Thus the liquid drop model predicts two possibilities: unstable nuclei that undergo fission, and spherical nuclei that do not.
This is illustrated in Fig.~\ref{fig:PES}.
However, the presence of shell structure changes the situation, and it is possible that the energy is minimized for a finite, non-zero deformation~\cite{BohrMottelson2,RingSchuck,Rowe2010}.
As with vibrations, the dominant multipole for deformations is quadrupole ($\lambda=2$), which can lead to shapes that are \keyword{prolate} (extended along $z$) or \keyword{oblate} (compressed along $z$).

A deformed nucleus breaks rotational symmetry; it has chosen a preferred direction (which by convention is defined to be the $z$ axis).
This \keyword{intrinsic state} should be projected onto good angular momentum $J$.
This amounts to rotating the deformed shape, in the same sense that an electron in an atom ``rotates'' around the nucleus.
This rotation comes with associated kinetic energy $E=\frac{J(J+1)}{2\mathcal{I}}$, where $\mathcal{I}$ is the moment of inertia of the deformed nucleus.
We therefore find a \keyword{rotational band} of states built from the same intrinsic state, rotating with different angular momentum $J$, analogous to the rotational bands found in molecules.


To illustrate how shell structure facilitates a deformed ground state, consider $^{41}$Sc ($Z\!=\!21$, $N\!=\!20$), modeled as a single proton added to a spherical, doubly-magic $^{40}$Ca core.
In the simple shell-model picture, the additional proton should go in the $0f_{7/2}$ orbit.
There are $2j+1=8$ degenerate states, differing in the projection $m_j$; we define the quantization axis so that $m_j=+j$.
In this case the projection of the orbital angular momentum is maximal $m_l=l$, so the spatial wave function lies predominantly in the $x\text{-}y$ plane; it is non-spherical.
Experimentally, $^{41}$Sc has an electric quadrupole moment of
$Q=-14.5(3)~e\,{\rm fm}^2$~\cite{Stone2013},
indicating an oblate shape consistent with the shell model picture.
Because the density is non-spherical, the corresponding mean field is non-spherical.
The nucleons in the $^{40}$Ca core could lower their energy by adjusting their orbits to better align with the oblate mean field, but this would amount to admixing orbits above the Fermi surface, which is inhibited by the shell gap.

Adding two neutrons to form $^{43}$Sc, the degeneracy of the $m_j$ orbits has been broken, and the neutron $0f_{7/2}$ orbits with $m_j=\pm j$ lie predominantly in the $x\text{-}y$ plane, maximizing alignment with the deformed mean field.
The neutrons carry no net electric charge, and so do not contribute directly to the electric quadrupole moment.
However, they enhance the deformation of the mean field, partially overcoming the shell gap so that the core orbits can adjust.
The electric quadrupole moment of $^{43}$Sc is $Q=-27(5)~e\,{\rm fm}^2$~\cite{Stone2013}, essentially double that of $^{41}$Sc.
This effect of the neutrons is an example of the phenomenon of \keyword{core polarization}.
For a detailed discussion in the context of the shell model, see~\cite{Brown1977}.


As described above, a deformed mean-field potential will in general alter the energies of single-particle orbits~\cite{Rainwater1950}.
The evolution of the spectrum of single-particle states with deformation can be studied with the \keyword{Nilsson model}~\cite{Nilsson1955}, in which the 3D harmonic oscillator potential is made anisotropic by taking a different oscillator frequency along the $z$ direction: $\omega_x=\omega_y \neq \omega_z$, parameterized by a deformation parameter $\delta$, so that the nuclear volume is conserved, $\omega_z^2/\omega_x^2 = (1-\tfrac{4}{3}\delta)/(1+\tfrac{2}{3}\delta)$.
For different deformations $\delta$, the gaps in the single-particle spectrum appear at different particle numbers, indicating that for certain combinations of protons and neutrons, a deformed shape can lower the energy~\cite{Rainwater1950,Nilsson1955}.
More details may be found in nearly any textbook on nuclear physics~\cite{BohrMottelson2,RingSchuck,deShalitFeshbach,WongIntro} and other contributions to this volume~\cite{AprahamianThis,CoelloThis}.

We may contrast this tendency for nuclei to adopt deformed shapes with the case of electrons in atoms, which also display shell structure, but which are generally spherical.
The essential difference is that the electron-electron interaction is repulsive, so that electrons maximize their separation, leading to Hund's rule.
Nuclei are deformed because their net residual interaction is attractive.
It is also necessary that the range of the interaction is short enough that there is preference among different orientations, but long enough that multiple nucleons can mutually interact.

When we say that a nucleus is deformed we are often referring to its shape in the intrinsic, or ``body-fixed'' frame in which the nucleus has a well-defined shape, but does not have a well-defined total angular momentum $J$.
Measurements are performed in the laboratory frame, where the nucleus has good $J$.
Inferring the intrinsic shape from experimental data is generally model-dependent~\cite{Poves2020}, because the probes usually correspond to one-body operators, while the intrinsic shape manifests as correlated motion of many particles.
In principle, relativistic heavy ion collisions can probe these many-body correlations directly~\cite{Jia2024}, though model-dependence reappears through the analysis of the collision~\cite{Dobaczewski2025}.




{\bf Competition between spherical and deformed shapes}

Whether a nucleus will be spherical or deformed in its ground state depends on the competition between pairing and deformation, and the underlying shell structure~\cite{Mottelson1960}.
Near the ``valley of stability'', a nucleus with a magic number of protons and neutrons will be spherical.
If only neutrons or only protons are added, pairing effects tend to win out, and nuclei are approximately spherical, with a superfluid character.
Examples include the calcium, tin, and lead isotopic chains.
With multiple protons and neutrons beyond a closed shell, static deformation eventually wins out.
Classic examples include the rare-earth elements between proton numbers 50 and 82 like gadolinium and dysprosium, and actinides beyond $Z=82$ like uranium and plutonium. 

This competition between shapes is evident even within a single nucleus, through a phenomenon called \keyword{shape coexistence}~\cite{HeydeRMP2011,PovesJPG2016}.
Many nuclei with a spherical ground state have a low-lying very deformed excited state, and vice versa.
For example, doubly-magic $^{40}$Ca has a spherical $0^+$ ground state, and a ``superdeformed'' $0^+$ first excited state at just 3.35 MeV~\cite{CaurierPRC2007}.
As protons are removed, increased correlations lower the energy of the deformed configuration until around $^{32}$Mg the deformed configuration becomes the ground state.
Regions of the nuclear chart in which this crossing occurs have been referred to as \keyword{islands of inversion}~\cite{WarburtonPRC1990,CaurierPRC2014}.

In principle, the wave function can include a superposition of spherical and deformed shapes that minimizes the energy, in a phenomenon called \keyword{shape mixing}~\cite{HeydeRMP2011,PovesJPG2016}.
Evaluating the ground state wave function in first order perturbation theory yields
\begin{equation} \label{eq:ShapeMixing}
    |\rm {g.s.}\rangle = |{\rm sph}\rangle +\frac{\langle {\rm def}| H | {\rm sph}\rangle}{\epsilon_{\rm sph}-\epsilon_{\rm def}}|{\rm def}\rangle 
\end{equation}
Where $|\rm{sph}\rangle$ and $|{\rm def}\rangle$ are the spherical and deformed configurations, with energies $\epsilon_{\rm sph}$ and $\epsilon_{\rm df}$, respectively.
If the spherical configuration is primarily given by the simple shell model configuration, and the deformed configuration corresponds to multiple particle-hole excitations out of that configuration, then the mixing will be small because the two-body interaction in $H$ can change the state of at most two particles.
Consequently, the two shapes will not strongly mix unless they come close in energy so that the denominator in \eqref{eq:ShapeMixing} becomes small.
One way to probe shape mixing is to study electric monopole ($E0$) transitions, which proceed primarily by pair production or conversion electrons~\cite{Wood1999}.

\subsection{The low-momentum limit: threshold effects and resonances\label{sec:WeakBinding}}

The shell model picture in section~\ref{fig:magicnumbers} was based on a harmonic oscillator potential $V\sim r^2 $, which grows to infinity as $r\to \infty$.
More realistically, the nuclear interaction is short-ranged, and so the potential should go to zero at large distance.
For well-bound states, the wave function does not probe the large separations where the oscillator potential becomes unphysical.
But for weakly-bound or unbound states, this difference becomes important.
The behavior near the threshold between bound and unbound depends on the charge and orbital angular momentum of the relevant single-particle state.

\begin{figure}
    \centering
    \includegraphics[width=0.45\linewidth]{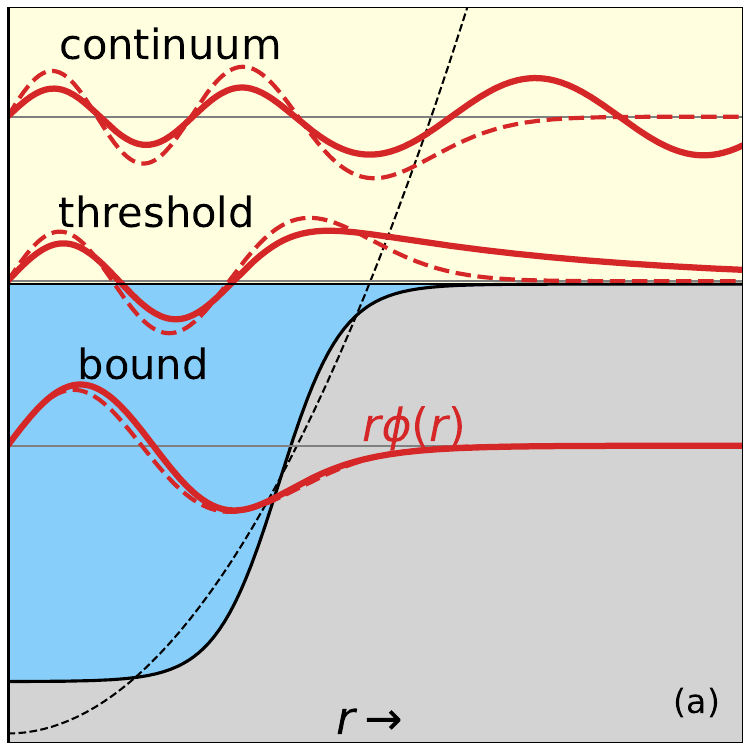}
    \quad
    \includegraphics[width=0.45\linewidth]{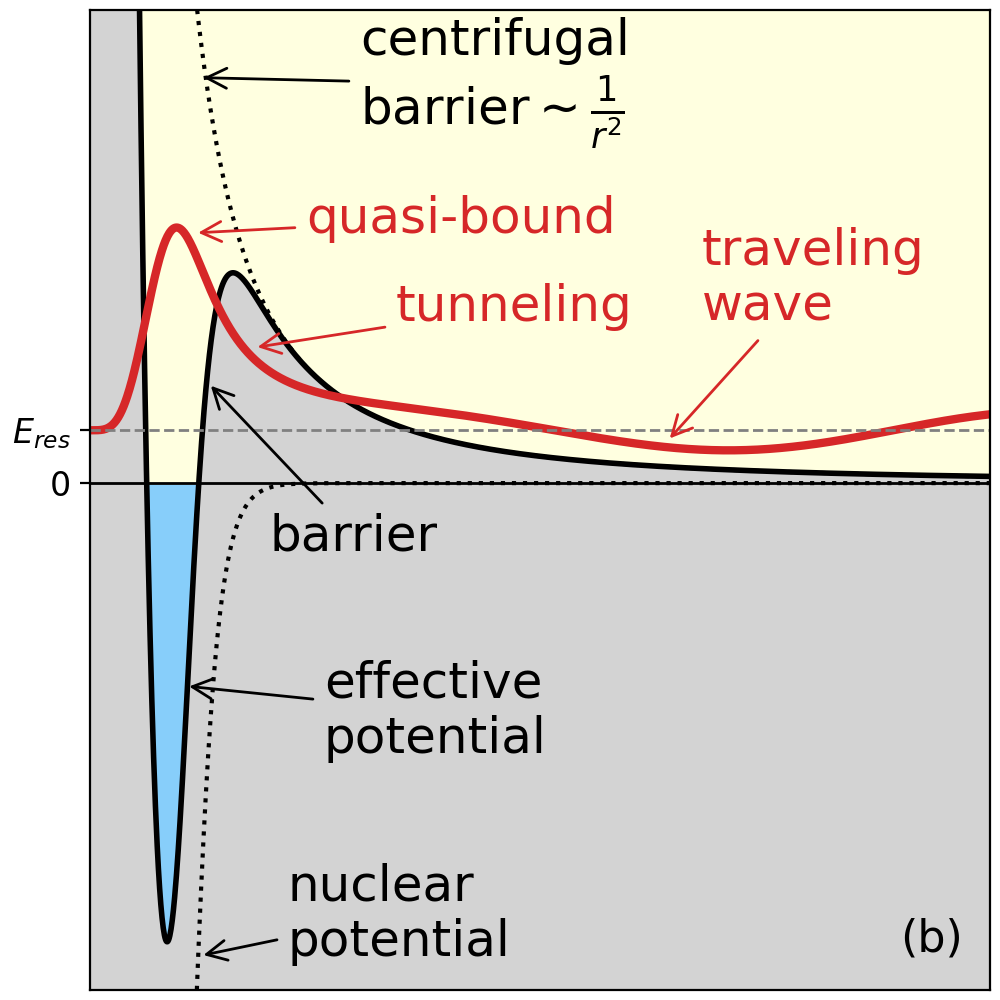}
    \caption{(a) Comparison of single-particle wave functions for a harmonic oscillator potential (dashed lines) and a Woods-Saxon potential which has bound and continuum solutions (solid lines) (b) Illustration of a resonance with an energy below the centrifugal barrier.}
    \label{fig:WScontinuum}
\end{figure}

For charged particles like protons or $\alpha$ partices, there is a repulsive Coulomb potential which extends beyond the range of the attractive nuclear potential.
Similarly, for both protons and neutrons, there is a centrifugal barrier $\hbar^2 l(l+1)/2mr^2$, which effectively acts as a potential for the radial motion.
These potentials create a barrier between the attractive interior $r\sim 0$ and the free exterior $r\to \infty$, as indicated in Fig.~\ref{fig:WScontinuum}(b).
When the energy of a single-particle orbit is above zero but below the height of the barrier, escape from the interior is hindered and we observe the phenomenon of a \keyword{resonance}~\cite{Taylor,KoenigThis}.
Resonances behave like quasi-bound states, but with an energy width $\Gamma\sim \tau/\hbar$, where $\tau$ is the decay time of the resonance.

For neutrons with $l=0$, there is no barrier, so resonances do not occur (though there is the related phenomenon of a \keyword{virtual state}~\cite{Newton,Taylor}).
Because there is no centrifugal barrier, weakly bound neutral $l=0$ states can extend into the classically-forbidden region while maintaining sufficient overlap with the attractive potential near $r\sim 0$.
As the binding energy approaches zero, the radial extent of the wave function grows, eventually yielding a \keyword{halo nucleus}, which can be viewed as a single nucleon in an extended cloud around a compact core.
This is illustrated in the curve labeled ``threshold'' in Fig.~\ref{fig:WScontinuum}(a).
In a halo nucleus a separation of scales emerges (e.g. the separation energy of the last neutron compared with the energy to excite the core), and an effective field theory can be developed, called halo effective field theory (or cluster effective field theory)~\cite{HammerRMP2020}.
In this regime, one finds nearly universal behavior, governed by a small number of parameters.
For more detailed discussion of halo nuclei, see~\cite{CapelThis}.

\subsection{\label{sec:SRC}The high-momentum limit: short-range correlations and spectroscopic factors}
As described in section~\ref{sec:ShellStructure}, the mean-field picture of the shell model appears to be in tension with the strong short-range repulsion illustrated in Fig~\ref{fig:AV18vsLJ}.
The argument that the Pauli principle blocks scattering is valid for energy transfers on the order of the fermi energy or less.
A strongly repulsive core can excite particles well above the fermi level, where Pauli blocking does not play a direct role.
The Pauli principle does play an indirect role, however.
The repulsive core strongly modifies the short-distance wave function between two nucleons, but modifications to the long-distance part of the wave function are still Pauli blocked, so that beyond a ``healing distance'' of $\sim 1~{\rm fm}$, the wave function of relative motion returns to its mean-field form~\cite{Gomes1958}.
Consequently, the mean-field picture is essentially accurate, except that two nucleons have a suppressed probability to be found close together (beyond the effects of the Pauli principle).
In this case we speak of \keyword{short-range correlations}.

One way to probe the presence of short-range correlations in the nucleus is with a direct reaction (see section~\ref{sec:DirectReactions}) that adds or removes a nucleon from a specific orbit, e.g. by inelastic electron scattering $(e,e'p)$, or transfer like $(d,p)$ (see \cite{PotelThis}).
The cross section for such a reaction can be (approximately) written in factorized form in terms of a single-particle cross section and a \keyword{spectroscopic factor}: $\sigma = \sigma_{\rm sp} S_{\alpha}$~\cite{AumannPPNP2021}.
The spectroscopic factor for orbit $\alpha$, defined as
\begin{equation}
    S_{\alpha} = \bigl|\langle \Psi_f^{(A)} | a^{\dagger}_{\alpha} | \Psi_i^{(A-1)}\rangle\bigr|^2,
\end{equation}
characterizes the extent to which the final state $\Psi_f^{(A)}$ consists of a nucleon in orbit $\alpha$ added to the initial state $\Psi_i^{(A-1)}$.
In the simple shell model picture, $S_\alpha=1$ for occupied shell model orbits and $S_\alpha=0$ for unoccupied orbits.
Any type of correlation will in general modify the spectroscopic factors from their simple shell-model values.
Long-range correlations like those discussed in section~\ref{sec:CollectiveModes} will in general depend on the low-lying level structure and will vary from nucleus to nucleus.
Conversely, the short-range correlation between a pair of nucleons will be largely independent of the environment, and so will manifest less system dependence.
More precisely, the probability to find two nucleons with relative separation $r$ and center of mass displacement $R$ factorizes in the limit $r\to 0$ as ~\cite{Cruz-Torres2020}
\begin{equation}\label{eq:GCF}
    \rho(r,R) \rightarrow C(R) \times |\varphi(r)|^2,
\end{equation}
where $\varphi(r)$ is a short-range function that is model-dependent but universal in the sense that it is nucleus-independent , while $C(R)$ (called the `nuclear contact coefficient') is model-independent but nucleus-dependent.

Systematic studies of various reactions on many nuclei conclude that nucleons in nuclei have an approximately 2/3 chance of being in their conventional shell-model orbits~\cite{PandharipandeRMP1997,KramerNPA2001}.
Further experiments have determined that the short-range correlations primarily consist of proton-neutron pairs with high relative momentum~\cite{HenRMP2017}.
One manifestation of this is that in reactions where a high-energy electron knocks a proton out of a nucleus, the proton is often accompanied by a high-momentum neutron.
This picture comes with the caveat that using a reaction to infer nuclear structure depends on how the reaction is modeled.

In 1983, the European Muon Collaboration published results~\cite{EMC1983} of deep inelastic scattering of high-energy muons on nuclei, essentially showing that the behavior of quarks within a nucleus is not simply given by $A$ copies of the quarks within a free nucleon.
Put another way, the nuclear environment changes the structure of the nucleon, with a magnitude that was unexpected based on the ratio of nuclear binding ($\sim 8$~MeV/$A$) to the energy transfer ($\sim$ several GeV).
Subsequent studies have shown that the effect grows with increasing mass number $A$.
Short-range correlations provide a possible explanation of this so-called ``\keyword{EMC effect}''~\cite{HenRMP2017}, though there is not yet a consensus in the field.
For a more detailed discussion of the EMC effect, see~\cite{ArringtonThis}.

\section{\label{sec:StructureToConstrainForces}Using nuclear structure to constrain the nuclear force}

Thus far, we have discussed how features of the nuclear force manifest as observed nuclear structure properties.
It is interesting to attempt to reverse the flow of information and see what nuclear structure observables tell us about nuclear forces.

Within a given theoretical framework, the most straightforward way to inform the nuclear force is to constrain its parameters.
(We focus here on chiral effective field theory, but most of the points carry over to other frameworks).
The two-body parameters are most naturally constrained in the two-body system, i.e. NN scattering and properties of the deuteron.
This is both because the system can relatively easily be solved exactly, and because a partial wave analysis of the NN scattering enables the disentangling of various components of the force.
Likewise, pion-nucleon coupling constants are most naturally and precisely constrained by pion-nucleon data~\cite{Hoferichter2015}.

It would be natural to constrain 3N forces in $A=3$ systems (in chiral effective field theory there are two unknown parameters in the leading 3N force).
However, the relevant data is highly correlated, so it does not sufficiently constrain the parameters of the 3N force.
In particular, isospin symmetry means that the $A=$ bound states, $^3$H and $^3$He, are essentially the same.
Two alternative choices, the neutron-deuteron scattering length~\cite{EpelbaumPRC2002} and the $^4$He binding energy~\cite{NoggaPRC2006} are highly correlated with the $A=3$ binding energy, reflecting the fact that at low energy, only one linear combination of the 3N terms contributes (this can be understood from the point of view that pionless effective field theory, with just one 3N parameter, is the relevant theory at low momenta)~\cite{HammerRMP2020}.
Other observables like the $^{4}$He charge radius~\cite{NavratilPRL2007} and the $^{3}$H beta decay half life~\cite{GazitPRL2009}, are also weakly constraining, due to correlations or cancellations~\cite{Jimenez2026}.

There are no bound $A=5$ systems, and even the $A=6,7$ binding energies do not appear to provide an independent constraint~\cite{LENPIC2019}.
One option is to include the binding energy and charge radius of $^{16}$O~\cite{EkstromPRC2015,JiangPRC2020,Huther2020,Arthuis2024}.
This approach tends to yield accurate binding energies up to heavy nuclei~\cite{JiangPRC2020,HuNP2022,Hu2025Texas}, but it has also been criticized for sacrificing reproduction of NN scattering~\cite{Machleidt2023}.
One challenge with using a many-body system like $^{16}$O to constrain nuclear forces is that all terms in the potential are active, and can be correlated in non-trivial ways.
One must explore a high-dimensional parameter space with relatively expensive many-body calculations.
Fortunately, the recent development of \keyword{emulators} has enabled such calculations in a statistically rigorous way~\cite{Frame2018,Ekstrom2019,Konig2020,HuNP2022}. 

Beyond fixing parameters, it is interesting to consider what general constraints nuclear structure data places on the form of the nuclear force.
For example, are there any properties of nuclei with $A\gg 2$ that reflect the broken chiral symmetry of quantum chromodynamics~\cite{FurnstahlJPG2010}?
Phenomenological mean-field theories accurately reproduce binding energies and radii for $A\gtrsim 16$ without explicitly including pion exchange, whose spin-isospin structure approximately averages to zero in bulk matter~\cite{Serot1992}, suggesting chiral symmetry is not essential for binding energies (see also~\cite{Lu2019}).
One possible positive example is the evolution of shell structure in neutron-rich nuclei, in particular effects which have been ascribed to the tensor force which arises primarily due to pion exchange~\cite{OtsukaPRL2005,OtsukaRMP2020}.
It has not yet been explored quantitatively the extent to which the observed shell evolution requires a light pion.

\begin{acknowledgments}
This work was supported by the NSF under Grant No. PHY-2340834, the NSF FRHTP PHY-2402275, and the DOE Topical Collaboration “Nuclear Theory for New Physics,” Award No. DE-SC0023663.
\end{acknowledgments}

\bibliography{References}

\end{document}